# A Generic Sharding Scheme for Blockchain Protocols

Research Thesis

## Zuphit Fidelman

This research was carried out under the supervision of Prof. Roy Friedman, in the Faculty of Computer Science.

The generous financial help of the Technion is gratefully acknowledged.

# Contents

# Abstract

Bitcoin was the first successful attempt at creating a decentralized cryptographic payment system. Overcoming the trusted third party hurdle, which plagued earlier solutions for decades, it paved the way for a whole family of blockchain protocols. However, its fully replicated design prevents Bitcoin from scaling. The scalability crisis is not unique to Bitcoin as this issue extends to most of its descendant blockchain protocols. A new paradigm that enables to systematically scale blockchain protocols seems due.

This thesis introduces a formal general framework for scaling blockchain protocols by sharding. The framework is modular and it can be adjusted for different needs or sets of assumptions. We prove that sharded protocols obtained by following our scheme (with correct modules in place) live up to the same safety and liveness guarantees as their non-sharded counterparts. The proof is general and relies on well-defined specifications of certain components. This lays the ground for simple proofs of correctness for sharded protocols obtained by following the proposed scheme.
 The framework is not left as an obscure specification of some high level structure; explicit use is demonstrated by applying it to shard Algorand. As part of this concrete construction, a tamper-proof mechanism to assign nodes to shards is introduced. This mechanism is constructed by using verifiable random functions and can safely withstand a powerful adaptive adversary.

# Abbreviations and Notations

| | | |
|---|---|---|
| $n$ | : | the number of users in the system |
| $m$ | : | the number of shards in the system |
| $r$ | : | an arbitrary round index |
| $TX$ | : | transaction space |
| $UTXO$ | : | unspent transactions output |
| $PKS$ | : | public-key space |
| $PK^r$ | : | the set of all public keys in the system at time $r$ |
| $C^r$ | : | the state of an unsharded system when round $r$ begins |
| $B^r$ | : | a block of transactions, approved by an unsharded system in round $r$ |
| $C_i^r$ | : | the state of the $i^{th}$ shard when round $r$ begins |
| $C^{(r)}$ | : | the state of a sharded system when round $r$ begins |
| $B_i^r$ | : | a sub-block of transactions, approved by the $i^{th}$ shard in round $r$ |
| $B^{(r)}$ | : | a global block of transactions, approved by a sharded system in round $r$ |
| $P$ | : | an arbitrary blockchain protocol |
| $[i]$ | : | $\{1, 2, \ldots, i\}$ |
| $\perp$ | : | the empty block |
| $\circ$ | : | concatenation (e.g. $a \circ b \circ c = abc$) |

# Chapter 1

# Introduction

## 1.1 Background and Motivation

Attempts to create cryptographic payment systems date back to the 1980s and 1990s. Those early endeavors relied on trusted third parties, deeming such solutions secure but centralized. In 2008 a whitepaper titled "Bitcoin: A Peer-to-Peer Electronic Cash System" [16] introduced the original Bitcoin protocol and coined the term *blockchain*. The paper was subsequently followed by the Bitcoin network's launch in early 2009. Bitcoin was the first truly decentralized electronic payment system, bypassing financial institutions, mediators and other trusted third parties. Nodes in the Bitcoin network store a full copy of a shared transactions ledger. Every new block of approved transactions is verified by all participating nodes, as they append it. Nodes compete over finding the next block, along with a solution to a cryptographic riddle. Overall, the collective computational power across the network takes the trusted third party's role.

This design marked a breakthrough, overcoming the trusted third party hurdle. It inspired the creation of an entire new category of blockchain protocols, modeled after Bitcoin. At the same time, this design prevents Bitcoin from scaling, marking its days as a global payment network numbered. The scalability crisis is not unique to Bitcoin, as this issue extends to most blockchain protocols that followed.

A new formal scheme that enables to systematically scale a whole category of blockchain protocols is the subject of this thesis.

**Blockchain Protocols and Their Usage**

Blockchain is a data model that keeps track of ordered sequences of transactions in a decentralized and persistent manner. It is known first and foremost as a platform for implementation of cryptocurrencies, such as Bitcoin [16], Ripple [17], Litecoin and others. Several other applications adopted this technology as well, most notably smart contracts [6] and supply chains [1, 12].

Different blockchain protocols propose different recipes on how to create an agreed upon shared transaction ledger, often under different sets of assumptions.

**Proof-of-Work and Proof-of-Stake** In order to withstand adversarial behavior, blockchain protocols employ various defense mechanisms, suited for a decentralized setting. Most notable are *proof-of-work* (PoW) and *proof-of-stake* (PoS) approaches. In PoW protocols, a user's influence on the system is proportional to its computing power. Such protocols provide safety as long as the majority of computing power in the network is held by honest participants. In PoS protocols, it is proportional to a user's stake (e.g. money) in the system and safety depends on the majority of overall wealth being held by honest parties.

**Permissioned and Permissionless** Loosely speaking, blockchain protocols can be categorized as *permissioned* vs. *permissionless*. In permissioned protocols, such as Ripple [17] and Hyperledger Fabric [2], only known nodes, who have permission, can take part in the system. In permissionless protocols, such as Bitcoin [16], Ethereum [6] and Algorand [7], any node may participate at will.

**Forks** Some blockchain protocols, such as Bitcoin [16] and Ethereum [6], may *fork*, i.e., there may temporarily exist multiple conflicting versions of the chain until the conflict is finally resolved. Other protocols, such as Algorand [7], *do not fork*, or may fork only with negligible probability.

## 1.2 The Scalability Debacle

In most existing blockchain protocols the system behaves as one fully replicated state machine, regardless of the network size. The state progresses by appending one block at a time to the end of the chain. This raises two main scalability concerns:

1. **Bounded throughput, waste of available computing power:**
   Once additional participants join the network, overall computing power increases and (presumably) more transactions pend approval. Assuming that the network has already been approving transactions at its maximum capacity, the added computing power could theoretically be used to serve those added transactions' approval. However, the intervals between block approvals must remain unchanged and still only a single block at a time, whose capacity is bounded by the protocol, could be appended to the chain. Overall, the network's throughput remains bounded by some fixed maximum capacity.

   This problem is most prominent with Bitcoin and similar PoW protocols. As the collective computing power increases, nodes have to work proportionally harder in order to simply maintain the same old throughput, let alone increase it.

2. **Waste of space:**
   The entire global state must be locally stored by all nodes that wish to participate in the protocol. In relatively small networks this indeed appears to be essential.

However, surpassing some desired redundancy threshold, this simply wastes space. Moreover, requiring nodes to store the global state, which could be sizable, in its entirety leaves lighter nodes with limited storage out of the game, thus impeding decentralization.

In a theoretical global network with millions of participating nodes, it seems absurd to store millions of identical copies of an entire enormous global state. Still, most current protocols are designed that way. This distances them from becoming those idealized decentralized networks used by the masses.

### 1.2.1 Sharding

Both scalability issues could be ameliorated through sharding.
*Sharding* is a design pattern for obtaining scalability of computing systems, by dividing the computational tasks and the data space into multiple entities, that act mostly independently. Sharding blockchain protocols intuitively means dividing nodes in a large network into smaller groups, where each group is in charge of approving a distinct subset of pending transactions and of storing a subset of the global state.

## 1.3 Contributions

This work presents a formal generic framework for sharding blockchain protocols.
The proposed framework is formal as it carefully defines all that must be defined and correctness is asserted thoroughly. The framework is generic as it follows a modular design which enables key components, including the blockchain protocol itself, to be easily swapped and independently created. The framework is general purposed and suits both the UTXO model and the account balance model. Moreover, it suits both permissioned and permissionless protocols and it mostly targets PoS protocols.

The framework executes in two-phased rounds, within a strongly synchronous network. During the first phase, disjoint sub-blocks of approved transactions are simultaneously committed by all shards. During the second phase, relevant cross-shard data is synchronized. Throughout this two-phased execution a specified workflow is followed, while maintaining three invariants. Each invariant is associated with a corresponding interface that defines a specific module. Applying the framework can be done simply by following our scheme while providing concrete implementations to all these interfaces. We provide a formal proof that as long as our scheme is correctly followed, i.e. all interface implementations maintain the desired invariants, then the new sharded protocol is indeed correct - satisfying safety and liveness. A useful result of this proof, whose only assumption is that supplied modules match their interface specifications, is that any subsequent protocols following our scheme, need only prove their modules' correctness to establish the overall correctness.

We demonstrate the applicability of our framework by sharding Algorand [7, 14]. A concrete implementation of all required interfaces is provided and the implementation's correctness is fully proved. Then, correctness of the sharded Algorand immediately follows. The methods used to implement the sharding interface are quite general. As we already established their correctness, these specific modules could be reused to shard protocols other that Algorand too, without having to re-establish their correctness.

**Resiliency** A major standing issue with sharded protocols is resiliency in presence of adversarial behavior. A potential attack vector by an adaptive adversary is to move all Byzantine nodes into a single shard, thereby controlling that shard despite controlling only a small minority of nodes overall.

The implementation presented in chapter 4 solves this issue by assigning participants time-bounded leases to uniformly selected shards. The assignment mechanism is decentralized and the generated assignments are both uniformly random and tamper-proof by using verifiable random functions (VRFs) to implement it. These nodes-into-shards assignments are iteratively shuffled, at a shuffling rate corresponding to the adversary's adaptability, as determined by the environment assumptions model. The provided proof establishes that as long as the adversary controls no more than $1/4$ of all nodes, then it never controls more than $1/3$ of the nodes in each shard with extremely high probability, even for extraordinarily long executions (e.g. the probability of ever encountering failure $< 10^{-15}$, totalled over a million years of continuous execution). Thus, the suggested implementation maintains safety despite Byzantine mobility.

## 1.4 Thesis Outline and Organization

The rest of this thesis is organized as follows:
In chapter 2 a formal model expressing the behavior of blockchain protocols is laid out. In chapter 3 a sharding scheme is introduced and its correctness is established, based on the definitions detailed in chapter 2. In chapter 4 a concrete construction implementing the sharding scheme is given and two sharded Algorand variants are created. In chapter 5 we survey recent work regarding sharding blockchain protocols. We also discuss the different approaches taken towards handling cross-shard transactions and compare them to the approach taken in this work. In chapter 6 we discuss the applicability of our framework and relaxation of some its assumptions.

# Chapter 2

# Preliminaries, Model and Definitions

We consider a distributed system in which *nodes* communicate through a strongly synchronous gossip network. Nodes are computational entities that execute some protocol. Nodes may join and leave the network, or the set of nodes could remain constant throughout the execution. Each *user* owns a pair of public and secret keys. When a user joins the system for the first time, their public key is recorded and this log is publicly available. Each node is owned and operated by a specific user. We use the terms "node" and "participant" interchangeably when referring to the machine used to participate in the protocol. By "user", we shall refer to the client operating the node and using the system (to send payments, for example).

The end goal of the system is to run some protocol, denoted $P$. In our case $P$ would be a blockchain protocol, that can be either permissioned or permissionless. $P$ is assumed to be a PoS protocol where users hold voting power in proportion to their stake in the system. Alternatively, if all nodes are guaranteed to possess equal computing powers, then $P$ can also be a PoW protocol. In order to simplify matters, from this point forward we regard all participants as equally powerful and assume that wealthier players have more distinct nodes under their control.

**Notations:** $n$ denotes the number of nodes, $\bot$ denotes the empty block, $\circ$ means concatenation (e.g. $a \circ b \circ c = abc$) and $[i] = \{1, 2, \ldots, i\}$.

## 2.1 Assumptions About the Environment

### 2.1.1 Users, Keys, Public-Key Space

Every *user* owns a pair of public and secret keys. Let $u$ denote some arbitrary user in the system, then $PK_u$ denotes $u$'s *public key* and $SK_u$ denotes $u$'s *secret key*. $PK_u$ is known to all participants and identifies $u$ uniquely. $SK_u$ is known only to $u$ and is used for authentication.

*PKS* denotes the *public-key space.* It is the set of all strings that could be used as acceptable public keys. For any $u$ in the system, it holds that $PK_u \in PKS$.

$PK^r$ refers to the set of all public keys in the system at time $r$.

### 2.1.2 Additional Cryptographic Assumptions

We assume the existence of a *unique digital signature scheme* `SIG`, as defined in [8, 11, 14]. `SIG` is used to generate $(PK_u, SK_u)$ pairs in the system. It provides $\texttt{Sign}_{SK_u}$ and $\texttt{Ver}_{PK_u}$ functionality. Specifically:

- `SIG` is a digital signature scheme:
    - $\texttt{Sign}_{SK_u}(x) \triangleq \texttt{Sign}(SK_u, x)$ returns a value $\sigma$.
    - $\texttt{Ver}_{PK_u}(x, \sigma) \triangleq \texttt{Ver}(PK_u, x, \sigma)$ returns *True* if $\sigma = \texttt{Sign}_{SK_u}(x)$.
      It is hard to find $\sigma$ for $x$ that returns *True* without knowing $SK_u$.
      It is hard to find $x' \neq x$ such that $\texttt{Ver}(PK_u, x, \sigma) = \texttt{Ver}(PK_u, x', \sigma) = $ *True*.

- `SIG` is unique:
    - It is hard to find $PK_u, x, \sigma$ and $\sigma' \neq \sigma$ such that:
      $\texttt{Ver}(PK_u, x, \sigma) = \texttt{Ver}(PK_u, x, \sigma') = $ *True*.
      This holds when $SK_u$ is known, even for maliciously selected $(PK_u, SK_u)$,
      $\sigma = \sigma_{PK_u}(x)$ is the unique signature for $x$ and $PK_u$.

We assume the existence of an efficiently computable ideal hash function, $H$.

- $H$ is available to all participants and they can run it locally.

- $H$ is modeled as a random oracle. It takes an arbitrarily long input and returns a *hashlen* bits long binary string. When $x$ is chosen uniformly at random:
    - $H(x)$ is regarded as a uniformly distributed integer between 0 and $2^{hashlen}-1$.
    - $.H(x)$ is regarded as a uniformly distributed binary between 0 and 1.

It should be pointed out that we adhere to the same cryptographic assumptions Micali et al. make in Algorand [7, 14], as we extend some of their constructions.

### 2.1.3 Adversarial Model

The adversary is a polynomially bounded entity, able to attack any target user's node and attain total control over their actions. It cannot successfully forge digital signatures by uncorrupted users and cannot simultaneously control or attack more than a fraction $0 \leq b < 1$ of active users; in chapter 4 we set $b = 1/4$.

A participant under adversarial control is considered *Byzantine*. A Byzantine participant may follow the protocol or it may deviate arbitrarily. All non-Byzantine participant are *honest* and meticulously follow the protocol. Gaining control over a set of target participants takes $t_{takeover}$ rounds to complete. During these rounds targeted nodes are still honest. Once an attack has been set in motion, it cannot be altered or aborted; the adversary must see it to completion.

We sometimes refer to an adversary as being either static or adaptive. A static adversary can introduce malicious participants to the network, but it can never gain control over honest ones. Equivalently, for a static adversary $t_{takeover} \to \infty$. An adaptive adversary is mobile, it can actively attack honest participants and gain control within a finite $t_{takeover} \in \mathbb{N}^+$ number of rounds. If an adaptive adversary is at capacity (simultaneously controls or attacks $b$ of all participants) it must retire control over a participant in order to initiate an attack on another.

## 2.2 Blockchain Protocols

We discuss blockchain protocols that execute in a sequence of consecutive rounds. Round $r \in \{1, 2, ...\}$ results in a block of approved transactions, denoted $B^r$, appended to the chain. The initial state of the system is determined by a genesis block, $B^0$.

Any $P$ we consider requires an honest majority of $0.5 < h_P \leq 1$ of all participants, i.e. $0 \leq b < 1 - h_P$. Then, it provides safety and liveness guarantees: all admitted blocks must be admissible (soon to be defined) and some of them non empty. We limit the discussion to protocols that **_do not fork_** [1]. When round $r$ completes, $B^r$ is finalized and its content is indisputable.

Let us introduce some definitions in order to dive into a more rigorous discussion:

**Definition 2.1** *Transaction Space:*
Let $P$ be some blockchain protocol. $TX_P$ denote the *transaction space* of $P$, or simply $TX$ since $P$ is known. A transaction is an atomic operation, defined per $P$'s logic. Any transaction $tx$ that could be created at some point in some execution of $P$ satisfies $tx \in TX$. Transactions contains various data fields, dictated by $P$. At the bare minimum, a $tx$ in all protocols we consider is assumed to include the following fields:

- *tx.from* : stores the identity of the user who created *tx*.
  This field must include exactly one public key, denoted $PK_{from} \in PKS$.
    - In a payment system $PK_{from}$ would be the public key of a user sending a payment, or transferring some resource it owns.

- *tx.to* : stores identity of the user who receives the output of *tx*.
  This field must include at least one $PK_{to} \in PKS$, more if *tx* has multiple outputs.
    - In a payment system $PK_{to}$ would be the public key of a user receiving a payment, or some other resource.

- *tx.sig* : stores $\texttt{Sign}_{SK_{from}}$ signature on this unique *tx*, used for authentication.

---
[1] This restriction could later be lifted, see discussion section. It simplifies matters for the time being.

– if $PK_{from}$ is a multisig address, this requires *m-of-n* signatures.

Any $tx'$ with either of these fields is missing or invalid satisfies $tx' \notin TX$. Note that a syntactically incorrect transaction is easily detectable and $tx'$ is immediately discarded.

**Definition 2.2** *Block:*
A *block* of transactions is any set $X$ such that $X \subseteq TX$. Usually (and unless stated otherwise), blocks are viewed as lists or ordered sets of transactions and then $X \in TX^{\mathbb{N}}$. Sometimes, blocks can be viewed as unordered sets, in which case $X \in 2^{TX}$.

**Definition 2.3** *State, Context:*
$C^r$ denotes the *context* of the protocol when round $r$ begins. The admissibility of any block to be added to the chain in round $r$ is determined by $P$ with respect to $C^r$.
The *state* of the system is defined by an ordered list of approved blocks. It is normally assumed that the context and the state equal, therefore we usually use $C^r$ to denote both. The state at time $r$ is recursively defined by $C^r = C^{r-1} \circ B^{r-1}$ with $C^0 = \bot$.

In settings where blocks are ordered sets, the state is *totally ordered*, $C^r \in (TX^{\mathbb{N}})^{\mathbb{N}^+}$. In settings where blocks are unordered sets, the state is *weakly ordered*, $C^r \in (2^{TX})^{\mathbb{N}^+}$. Explicitly: take some $tx_i \in B^i$, $tx_j \in B^j$. If $i < j$ then $tx_i$'s approval *predates* $tx_j$'s. If $i = j$ and blocks are ordered sets, then $tx_i, tx_j$ approval order is induced by the internal order in that block. If $i = j$ and blocks are unordered sets, then $tx_i, tx_j$ are considered to be simultaneously approved.

We refer to $B^0$ as the genesis block and to $C^1 = \bot \circ B^0$ as the genesis state, both are assumed to be valid. Applying an admissible block on a valid state transitions the system into a new valid state. This leads us to the next definition.

**Definition 2.4** *Admissible Blocks, Valid States, Legal Executions:*
A block is *admissible* at time $r$ if all transactions in that block, in order of appearance, could be appended to the chain and admitted to the system, in compliance with the protocol's logic. $P$ is assumed to include a well defined deterministic function named *Verify*. It takes an ordered set of transactions along with a context $C^r$ and returns *True* if the set is admissible with regards to $C^r$ or *False* otherwise. Formally:

$$\textit{Verify} : (TX^{\mathbb{N}}, (TX^{\mathbb{N}})^{\mathbb{N}^+}) \to \{\textit{True}, \textit{False}\}$$

$C^0 = \bot$ is a *valid state*, the rest is inductively defined such that:

$$\textit{Verify}(B^r, C^r) = \textit{True} \iff C^{r+1} = C^r \circ B^r \text{ is } \textit{valid} \text{ , } B^r \text{ is } \textit{admissible} \text{ at time } r$$

An execution is *legal* if and only if all of its states are valid.

$P$ must never permit invalid intermediate states. Any prefix of an admissible block must also be admissible on its own, with respect to the same state. Take $B^r$ such that $\textit{Verify}(B^r, C^r) = \textit{True}$ and any prefix of $B^r$, denoted $X$. Then, $\textit{Verify}(X, C^r) = \textit{True}$ always holds. Additionally, $\textit{Verify}(B^r \setminus X, C^r \circ X) = \textit{True}$ should also hold.

*Remark.* In practice, blocks contain additional bookkeeping records (such as headers, PoW etc.) on top of the list of approved transactions. However, our interest lies in the state of the ledger, as determined by those ordered sets of transactions. Therefore, in order to keep this focus, we abstract. Any additional bookkeeping that might be required throughout this work will be handled separately and explicitly. Alternatively, consider *Verify* as one component of a more comprehensive *Validate* method.

**Definition 2.5** *Intra-Block Transactions:*
$tx \in B^r$ is considered an *intra-block transaction* if its input depends on the output of another transaction $tx' \in B^r$ preceding it within the same block: $B^r = [\ldots, tx', \ldots, tx, \ldots]$ such that $Verify(B^r, C^r) = $ *True* and $Verify(B^r \setminus \{tx'\}, C^r) = $ *False*.

**Block Permutation Effect:** The existence of intrablock transactions gives rise to cases where two blocks, $B^r$ and $B^r_\pi$, consist of equal sets of transactions in different orderings, such that $Verify(B^r, C^r) = $ *True* while $Verify(B^r_\pi, C^r) = $ *False*. Take, for example, $B^r = [\ldots, tx', \ldots, tx, \ldots]$ from definition 2.5 such that $Verify(B^r, C^r) = $ *True* and $Verify(B^r \setminus \{tx'\}, C^r) = $ *False*. Define $B^r_\pi = [\ldots, \ldots, tx, \ldots, tx'] = [B^r \setminus \{tx'\}] \circ tx'$. $Verify(B^r_\pi, C^r) = $ *False* follows from definition 2.4, as states are totally ordered.

Such cases may cause ambiguity and raise additional difficulties when parallelizing *Verify* or transforming a system that outputs one fully replicated block at a time into one producing several sub-blocks, separately and simultaneously. For these reasons, from this point forward, we consider only protocols (or variants of protocols) that *do not permit intrablock transactions*. Specifically, we require *Verify* to always return $Verify(B^r, C^r) = Verify(B^r_\pi, C^r)$ for all $B^r, C^r$ and any possible permutation $\pi$.

**Lemma 2.2.1.** *If $P$ does not permit intra-block transactions, then any subset $X \subseteq B^r$ of an admissible $B^r$ also makes an admissible block, applicable instead of $B^r$ at time $r$. Formally: $Verify(B^r, C^r) = $ True $\Rightarrow \forall X \subseteq B^r : Verify(X, C^r) = $ True.*

*Proof.* For any permutation $\pi$ it holds that $Verify(B^r, C^r) = Verify(B^r_\pi, C^r) = $ *True*. Consider a permutation $\pi$ where $X$ appears as the first $|X|$ transaction within $B^r_\pi$, in some internal order. Any prefix of $B^r_\pi$, in particular one of length $|X|$, must also be admissible. □

***We do not permit intra-block transactions***, i.e. we do not allow a resource received within the round to be immediately spent.[2] All resources being put to use in round $r$ must already be admitted when the round began, appearing in $C^r$. Then, the internal order of $B^r$ plays no role. Funds or resources received during round $r$ become available to spend starting round $r+1$, as they appear in $C^{r+1}$. This strategy appears reasonable when rounds are frequent enough, it resolves such ambiguity altogether.

---

[2] Removing this assumption, sharded protocols created by the framework still maintain safety. See discussion section for more details.

From this point forward, as internal order is insignificant, we shall treat each block as an unordered set. All transactions appearing within the same block are considered simultaneously approved and *Verify* takes the form:

$$Verify: (2^{TX}, (2^{TX})^{\mathbb{N}^+}) \to \{\textit{True}, \textit{False}\}.$$

**Definition 2.6** *Conflicting Sets, Competing Transactions:*
Let $tx_1, tx_2 \in TX$ be two distinct transactions. If there exists some $X \subseteq TX$ and a state $C \subseteq (2^{TX})^{\mathbb{N}^+}$ such that $Verify(X \cup \{tx_1\}, C) = \textit{True}$ and $Verify(X \cup \{tx_2\}, C) = \textit{True}$ but $Verify(X \cup \{tx_1, tx_2\}, C) = \textit{False}$ then $tx_1, tx_2$ are **competing transactions** and $X \cup \{tx_1, tx_2\}$ is a **conflicting set** (note that this definition generalizes the notion of double spending [3]).

Let $X = \{tx_1, tx_2, ..., tx_k\}$ be a conflicting set s.t. $\forall tx_i \in X : X \setminus \{tx_i\}$ is an admissible set, then $X$ is **minimal**. Observe that all distinct pairs $tx_i, tx_j \in X$ are *competing transactions*. [4]

**Definition 2.7** *Support Sets:*
The *support set* of $X \subseteq TX$ at time $r$ is the minimal subset of $C^r$ that is required to correctly determine whether any $X' \subseteq X$ could be accepted or must be rejected, per $P$'s logic. Denoted $Supp_r(X) \triangleq Supp(X, C^r) \subseteq C^r$, this is a weakly ordered set. The order between elements is induced by their order in $C^r$.

$$Supp(X, C^r) \text{ satisfies } \quad \forall X' \subseteq X : Verify(X', C^r) = Verify(X', Supp(X, C^r))$$

Observe that:

1. $\forall X' \subseteq X : Supp_r(X') \subseteq Supp_r(X)$
2. $\forall C^r \supseteq Y \supseteq Supp_r(X) : Verify(X, Supp_r(X)) = Verify(X, Y)$

are both implied.

Note that *Supp* definitions are modeled after and rely on internal $P$ behavior. Since minimality is a requirement and definitions are $P$-specific, *Supp* is somewhat illusive. Therefore, we also define $Supp(X) \triangleq Supp(X, TX) \subseteq TX$ which splits $TX$ into transactions that may appear in $X$'s support and those that never will. Then, $Supp(X, C^r) \subseteq Supp(X) \cap C^r \subseteq C^r$ always holds and is easier to use.

---

[3] Under the UTXO model, if $tx_1, tx_2$ compete over the same *utxo*, they certainly conflict. Take, for example, $X = \emptyset$ then $X \cup \{tx_1, tx_2\}$ is a conflicting set (it is also minimal). Generally, a resource may be used or accessed more than once. Consider a venue that can hold up to 200 distinct visitors at a time. Visitors compete over the same resource (entrance to the venue), yet as long as fewer than 201 of them coexist, no conflict arises.

[4] Let $X$ be a set of 201 distinct visitors asking to enter an empty venue that is capped at 200. Then $X$ is a conflicting set, any proper subset of $X$ (in particular one of size 200) is admissible and therefore $X$ is minimal. The same analysis applies to the sum of monetary transactions under the account balance model. Building on top of these definitions makes our scheme general enough to be useful under both the UTXO and account balance models.

## 2.3  Sharded Blockchain Protocols

When dealing with sharded protocols, we slightly adjust some terminology. Let $S$ denote a sharded blockchain protocol and $m$ denote the number of shards. Then $S$'s participants are scattered between those $m$ shards. Nodes that participate in the $i^{th}$ shard run some intra-shard sub-protocol, denoted $S_i$, that sequentially outputs blocks of the form $B_i^1, B_i^2, \ldots, B_i^r, \ldots$. We generally assume that $S$ is synchronized. Specifically, if $S_i$ hasn't decided on $B_i^r$ within the $r^{th}$ time frame, then $B_i^r = \bot$ .

**Definition 2.8** *Global Blocks and Local Blocks:*
$B^{(r)}$ denotes the $r^{th}$ *global block*. $B^{(r)}$ is a distributed block rather than a fully replicated one. We define $B^{(r)} = B_1^r \circ \ldots \circ B_m^r$ or simply $B^{(r)} = \bigcup_{i=1}^{m} B_i^r$, since we consider $B^{(r)}$ to be simultaneously approved. We refer to $B_1^r, \ldots, B_m^r$ as *sub-blocks* or *local blocks*. Normally, $B_i^r$ is locally stored within $S_i$'s nodes. The sharded protocol $S$ is assumed to allow its participants to query any $S_j$ regarding the content of any $B_j^r$.

**Definition 2.9** *Global State and Local Contexts:*
Given an environment in which the system is not fully replicated, some participants view of it may differ from others. This requires us to make a clear distinction between contexts and the state of the system.

$C^{(r)}$ denotes the *global state* when round $r$ begins. The global state of a sharded system follows the same definition a state in an unsharded system follows. Specifically, it is recursively defined by $C^{(r)} = C^{(r-1)} \circ B^{(r-1)}$ with $C^{(0)} = \bot$. The global state is the full ledger, defined by an ordered list of approved global blocks.

$C_i^r$ denotes the *local context* (sometimes called local state) of the $i^{th}$ shard at time $r$. This is some subset of the global state $C^{(r)}$, stored in $S_i$ nodes. Its participants execute $S_i$ and produce $B_i^r$ under that local context $C_i^r$. We specifically require $Verify(B_i^r, C_i^r) = True$ to hold, the rest of $C_i^r$ definition is dependant on the protocol $S$ at hand.

We use this different notation so to clearly distinct between sharded and unsharded systems. Any definition relating to the state (e.g. support sets) remains unchanged in the sharded setting, other than using $C^{(r)}$ instead of $C^r$.

# Chapter 3

# Shardder: The Sharding Framework

## 3.1 Overview

This chapter presents Shardder, a framework for transforming an overcrowded network running a single instance of $P$ into a network that runs several specialized sub-instances, working together in cooperation. These sub-protocols, denoted $P_1, \cdots, P_m$, are simultaneously executed and produce $m$ disjoint sub-blocks in parallel. The overall capacity of the network increases by up to $m$-fold, while preserving $P$'s original logic (as defined by its *Verify* predicate). Following a careful division of labor, $P_i$ executes correctly from both local and global perspectives. This is achieved despite the fact that participants store only a fraction of the global state locally.

Shardder is generic in a sense that while the framework follows a preset workflow, several key components are black-boxed into well defined interfaces. Each interface lays out a set of requirements that any adequate implementation must fulfill. The framework-user, i.e. the *deployer*, must provide implementations of three interfaces, a concrete protocol $P$ and a granularity parameter $m \in \mathbb{N}^+$. Plugging these components into the framework results in defining $SP$, a Sharded $P$ protocol. Swapping some interface implementation results in creating a different sharded variant of $P$. Different variants might be better suited for different settings, as demonstrated in chapter 4. Any sharded variant created by applying the framework lives up to original protocol's safety and liveness guarantees, provided that all interface implementations (provably) meet all specified requirements, this we prove in section 3.4.

### 3.1.1 Assumptions

$P$ is assumed to be some blockchain protocol that:

1. Does not permit intra-block transactions.

2. Does not fork. Block approval is final and cannot be overturned.

3. Provides safety and liveness, as long as at least $0.5 < h_P \leq 1$ of its participants are honest.

See discussion in section 6.1 regarding relaxing assumptions 1 and 2.

### 3.1.2 Key Parameters

$n$ - denotes the initial number of participants.

$m$ - denotes the number of shards set by the deployer. $m \in \mathbb{N}^+$ s.t. $n \gg m$.

$h$ - denotes the honesty rate the resulting $SP$ assumes. It must hold that $h_P \leq h \leq 1$. The precise value of $h$ is set by the deployer, in accordance with their concrete interface implementations, acceptable values for $m$ and adversarial assumptions.

We assume that the number of participants remains $\theta(n)$ throughout the execution and doesn't fluctuate too wildly. Permissionless environments in which specific users may join or leave are suitable, as long as the number of active users remains stable. [1]

### 3.1.3 Key Components

**Partition** $TX$ is partitioned into $TX_1, \ldots, TX_m$ according to the `Partition` interface. By definition, the sole authority regarding approval of pending transactions from within $TX_i$ is $P_i$.

**Membership** Participants are divided into $m$ groups by the `Membership` interface. It must ensure that at least $h_P$ out of every $P_i$'s participants are honest, thus enabling them to execute the underlying $P$.

**Sync** Global changes are tracked and local states are updated, using the `Sync` interface.

---

[1] Adjusting the framework for networks with significant fluctuations requires to dynamically adjust $m$ for $n/m$ as $n$ deviates. This extension is left for future work, see discussion section.

## 3.2 Algorithms and Workflow

### 3.2.1 Bootstrapping the System

To construct the system, concrete implementations of all abstractions used by the framework are required. This includes the code to a blockchain protocol $P$ and to non-abstract classes implementing the sharding interfaces. Additionally, genesis data is needed to set the genesis state. The framework takes a standard valid $P$-formatted genesis block, $B^0$, for this purpose.

During bootstrap, $m$ specialized instances of $P$, referred to by $P_1, \ldots, P_m$, are created. The instances are then initialized so to reflect the appropriate genesis state: $B^0$ data is broken into $m$ shards, according to `Partition`. Thus, transforming $B^0$ into an equivalent distributed block $B^{(0)}$ s.t. $tx \in B^0 \Leftrightarrow tx \in B^{(0)}$.

---

**Algorithm 1:** `Bootstrap` Shardder

**Input** : $P, B^0$, nodes, $m \in \mathbb{N}^+$, `Partition`, `Membership`, `Sync`
**Output :** Bootstrapped System

1 define $\mathsf{Part}_m = \mathtt{Partition.Part}(TX, m) = \{TX_1, ..., TX_m\}$;
2 init `Membership` and `Sync`;
3 **forall** $i \in [m]$ **do**
4      $C_i^0 = \bot$;
5      $B_i^0 = B^0 \cap TX_i$;
6      $RS_i^0 = \mathtt{Sync}.\,collectSupport_i(B^0)$ ;
7      construct $P_i$ and set $C_i^1 = C_i^0 \circ B_i^0 \circ RS_i^0$;
8 **end**
9 scatter nodes between $P_1, \ldots, P_m$ according to `Membership`;
10 **return** $P_1, \ldots, P_m$, $\mathsf{Part}_m$, $B^{(0)} = \bigcup_{i=1}^m B_i^0$;

---

### 3.2.2 Making Progress

After bootstrap, the framework follows Algorithm 2 and executes in sequential two phased rounds. Every round yields a new global block of approved transactions, which transitions the system into its next state.

During the first phase, the $P$ phase, $P_1, \ldots, P_m$ simultaneously execute a single block creation round. Every $P_i$ is essentially an instance of $P$ with protection against unauthorized participants, as described in Algorithm 3. Each $P_i$ executes under its local context $C_i^r$ and takes as input pending transactions from $TX_i$, its designated partition. Consequently, $P_i$ outputs a locally admissible block $B_i^r \subseteq TX_i$ such that $Verify(B_i^r, C_i^r) = True$. [2]

---

[2] An empty sub-block is decided by default if $P_i$ times out. Empty blocks are always admissible.

Reaching the second phase, the sync phase, $B^{(r)}$ is globally known and the next global state $C^{(r+1)} = C^{(r)} \circ B^{(r)}$ is determined. Then, during the sync phase and before transitioning to the next round, shards utilize the `Sync` interface in order to update their local states to include the up-to-date $Supp(TX_i, C^{(r)} \circ B^{(r)})$. [3]

---

**Algorithm 2:** Shardder Workflow

**Input** : a stream of pending transactions
**Output** : $B^{(1)}, B^{(2)}, ...$

**1** Bootstrap; $r = 1$;
**2 while** *safety holds* **do**
    // protocol phase
**3**    execute $P_1, \ldots, P_m$ on $TX_1, \ldots, TX_m$ respectively;
**4**    set $B^{(r)} = \bigcup_{i=1}^{m} B_i^r$ and make public;
    // sync phase
**5**    **forall** $i \in [m]$ **do**
**6**        $RS_i^r = \texttt{Sync}.collectSupport_i(B^{(r)})$;
**7**        $C_i^{r+1} = C_i^r \circ B_i^r \circ RS_i^r$;
**8**    **end**
**9**    adjust nodes' memberships;
**10**   $r = r + 1$;
**11 end**

---

**Algorithm 3:** $P_i$ block creation round, as executed by $PK_u$

**Input** : a set of pending transactions $T \subseteq TX_i$
**Output** : an approved block $B_i^r \subseteq T$

**1** let $\sigma_u$ be a proof binding $PK_u$ and $i$, generated by `Membership`.*getMembership*;
**2 while** *running round $r$ of $P$ with context $C_i^r$* **do**
**3**    attach $(PK_u, \sigma_u, i, r)$ to all messages directed to the underlying $P$;
**4**    **foreach** *received protocol message from node $w$* **do**
**5**        **if** `Membership`.*verifyMember*$(PK_w, \sigma_w, i, r) = $ *False* **then**
**6**            discard message;
**7**        **else** pass message to underlying $P$;
**8**    **end**
**9 end**
**10 return** $B_i^r$;

---

[3] Take notice of the fact that $P_i$'s synchronization with $B^{(r)}$ does not imply that every node must run or acquire every transaction. While this is a technical design issue, sub-blocks could batch together transactions that belong in distinct shards' support. Then, nodes need to download nothing more than the relevant sections of each sub-block.

## 3.3 Interfaces, Methods and Invariants

The sharded protocol's correctness relies on three invariants being kept throughout its execution. In order to plug each invariant into the framework, an interface abstraction is used. The interfaces specify a set of methods that must be implemented, as those methods are invoked by the framework. As long as concrete implementations supplied by the deployer satisfy the following requirements, $SP$'s correctness immediately follows from the proof in Section 3.4.

### 3.3.1 Conflict Preserving Partitioning

**Interface Name:** `Partition`

**Role:** Splits the workload into well defined pools, each allotted to a specific shard.

**Key Methods:**

1. $\mathsf{Part} : (2^{TX}, \mathbb{N}^+) \to (2^{TX})^{\mathbb{N}^+}$

   $\mathsf{Part}_m(TX) = \mathsf{Part}(TX, m) = \{TX_1, ..., TX_m\}$ partitions $TX$ into $m$ subsets.

2. $whichPart : TX \to \mathbb{N}^+$

   A companion function to $\mathsf{Part}_m$ s.t. $\forall tx \in TX_i : whichPart_m(tx) = i$.

**Invariant :**

1. **Partition Function:**
   $\mathsf{Part}$ is a partition function. $\forall m \in \mathbb{N}^+$ it must satisfy:
   $\mathsf{Part}(TX, m) = \{TX_1, ..., TX_m\}$ such that $TX = \bigcup_{i=1}^{m} TX_i$ , $\forall i \in [m] : TX_i \neq \emptyset$ and $\forall i \neq j : TX_i \cap TX_j = \emptyset$.

2. **Conflict Preservation / Aggregation:**
   $\mathsf{Part}$ must assign any pair of competing transactions into the same partition.
   $tx, tx' \in TX$ are competing transactions $\Rightarrow whichPart_m(tx) = whichPart_m(tx')$.

**Definition 3.1** *Remote Support:*
The remote support of $TX_i$ in $B^{(r)}$ is $RemoteSupp_i(B^{(r)}) \triangleq \{Supp(TX_i) \cap B^{(r)}\} \setminus B_i^r$.

### 3.3.2 Self Containment

**Interface Name:** `Sync`

**Role:** Enables participants to update local states so to reflect global changes.

**Key Method:**

1. $collectSupport : (2^{TX}, \mathbb{N}^+) \to 2^{TX}$
   $collectSupport_i(B^{(r)}) = collectSupport(B^{(r)}, i)$ return $RS_i^r \subseteq \{B^{(r)} \setminus B_i^r\}$
   such that $RemoteSupp_i(B^{(r)}) \subseteq RS_i^r$ always holds.

**Invariant :** Self Containment must hold throughout the first phase of every round.
*Self Containment* holds $\Leftrightarrow \forall i \in [m] : Supp(TX_i, C^{(r)}) \subseteq C_i^r$.

**Theorem 3.1.** *If a collectSupport function that returns $RS_i^r$ such that $\forall i \in [m], r \in \mathbb{N}$ : $tx \in RemoteSupp_i(B^{(r)}) \Rightarrow tx \in RS_i^r$ is used, then self containment holds throughout the execution.*

*Proof.* By induction:

**Base:** $C^{(0)} = C_i^0 = \bot$ .

**Closure ($r > 0$):** consider progress made by the system:
$C_i^r = C_i^{r-1} \circ B_i^{r-1} \circ RS_i^{r-1}$ as determined by Algorithm 2.
$C^{(r)} = C^{(r-1)} \circ B^{(r-1)}$ as defined in 2.9.

Take any $x \in Supp(TX_i, C^{(r)})$. Either $x \in C^{(r-1)}$ or $x \in B^{(r-1)}$.

- If $x \in C^{(r-1)}$ then $x \in Supp(TX_i, C^{(r-1)})$.
  By induction hypothesis, $Supp(TX_i, C^{(r-1)}) \subseteq C_i^{r-1}$ and we get $x \in C_i^{r-1}$.

- When $x \in B^{(r-1)}$ we have two cases to consider:
  - if $x \in B^{(r-1)} \cap TX_i$ then $x \in B_i^{r-1}$.
  - otherwise, $x \in \{B^{(r-1)} \setminus B_i^{r-1}\}$ meaning $x \in RemoteSupp_i(B^{(r-1)}) \subseteq RS_i^{r-1}$

Overall, $x \in C_i^{r-1} \circ B_i^{r-1} \circ RS_i^{r-1} = C_i^r \Rightarrow Supp(TX_i, C^{(r)}) \subseteq C_i^r$ □

### 3.3.3 Honest Majority

---
**Interface Name:** `Membership`

**Role:** Assigns nodes to shards in a tamper-proof manner.

**Key Methods:**

1. $getMembership : ((PK, SK), N^+, ...) \to (N^+, \{0, 1\}^*)$

   $getMembership((PK_u, SK_u), r, \ldots) \to \langle i, \sigma_u \rangle$ where $i \in [m]$.
   generates $\sigma_u$ that uniquely binds $PK_u$ with $P_i$

2. $verifyMember : (PK, \{0, 1\}^*, N^+, N^+, ...) \to \{True, False\}$

   $verifyMember(PK_u, \sigma_u, i, r) \to \begin{cases} True & getMembership((PK_u, SK_u), r) = \langle i, \sigma_u \rangle \\ False & \text{otherwise} \end{cases}$

   verifies $\sigma_u$ authenticity, for $PK_u \in PKS$, $i \in [m]$, $r \in \mathbb{N}^+$ such that
   any $\sigma_u'$ not issued by *getMembership* returns *False* with extremely high probability.

---

**Definition 3.2** *$PK_i^r -$ the set of $P_i$ participants at time $r$:*
$PK_i^r = \{PK_u \in PK^r \mid verifyMember(PK_u, \sigma_u, i, r) = True\}$

**Invariant :** Honest majority must hold at all times with extremely high probability.
honest majority holds $\Leftrightarrow \forall i \in [m]$ : at least $h_P \cdot |PK_i^r|$ of $P_i$'s participants are honest.

## 3.4 Correctness

In this section we prove that any $SP$ created by properly following the algorithms presented in section 3.2 is indeed correct. Namely, $SP$ satisfies the same safety and liveness guarantees that an instance of $P$ would have.

$SP$ is assumed to be correctly deployed, with all interface implementations satisfying the requirements listed in section 3.3. We examine the course of execution in two parallel $P$ and $SP$ networks that are executing under equal terms. Specifically, both networks are bootstrapped with the same genesis block, $B^0$, and are running with $n$ participants, out of which at least $h \cdot n$ are honest. Block-size limitations are disregarded and comparing $SP$ with an unbounded version of $P$, we demonstrate equivalent safety and liveness. Recall that:

- An execution $e$ of $P$ is legal $\Leftrightarrow$
  every $B^r$ outputted by $e$ satisfies $Verify(B^r, C^r) = True$.

- An execution $e'$ of $SP$ is legal $\Leftrightarrow$
  every $B_i^r$ outputted by $e'$ satisfies $Verify(B_i^r, C_i^r) = True$.

**Lemma 3.4.1** ($P \subseteq SP$). *For any legal execution $e$ of $P$ that runs with input $T \subseteq TX$ and outputs $B^0, B^1, B^2, ...$, there exists an equivalent execution $e'$ of $SP$ that runs for the same number of rounds, with the same input $T$ and outputs $B^{(0)}, B^{(1)}, B^{(2)}, ...$ s.t. $\forall r \geq 0 : B^{(r)} = B^r$.*

*Proof Sketch.* Given $e$ executed by $P$, we show the existence of a legal execution of $SP$, denoted $e'$, that always decides $B^{(r)} = B^r$. We inductively construct $e'$, round by round.

$r = 0$ : Let's explicitly demonstrate that Algorithm 1 always returns with $B^{(0)} = B^0$: This is indeed the case because $\forall i \in [m] : B_i^0 = B^0 \cap TX_i$, $B^{(0)} = \bigcup_{i \in [m]} B_i^0$ and since $\mathsf{Part}_m$ is a *partition function* $B^0 = \bigcup_{i \in [m]} B^0 \cap TX_i = \bigcup_{i \in [m]} B_i^0 = B^{(0)}$.

$r > 0$ : Take the inductively constructed $e'$ that agrees with $e$ up to round $r - 1$. Having $B^{(0)} = B^0, \ldots, B^{(r-1)} = B^{r-1} \Rightarrow C^{(r)} = C^r$ we need only to extend $e'$ by one additional round where $B^{(r)} = B^r$. $P$ does not allow intra-block transactions, therefore:

$$Verify(B^r, C^r) = True \Rightarrow \forall X \subseteq B^r : Verify(X, C^r) = True$$

The LHS holds since $e$ is a legal execution. Then the RHS holds too, it holds in particular for $\forall i \in [m] : X = B^r \cap TX_i$. Clearly $B^r \cap TX_i \subseteq TX_i$, we have established that $C^{(r)} = C^r$ and since *self containment* holds $Supp(TX_i, C^r) \subseteq C_i^r$. This gives $Verify(B^r \cap TX_i, C_i^r) = Verify(B^r \cap TX_i, C^r) = True$, which makes $B_i^r = B^r \cap TX_i$ a legal $P_i$ output for round $r$, for *every* $P_i$. Overall $B^{(r)} = \bigcup_{i \in [m]} B_i^r = \bigcup_{i \in [m]} B^r \cap TX_i = B^r$ makes legal round $r$ output in $e'$.

**Lemma 3.4.2** ($SP \subseteq P$). *For any legal execution $e'$ of $SP$ that runs with input $T \subseteq TX$ and outputs $B^{(0)}, B^{(1)}, B^{(2)}, \ldots$, there exists an equivalent execution $e$ of $P$ that runs for the same number of rounds, with the same input $T$ and outputs $B^0, B^1, B^2, \ldots$ s.t. $\forall r \geq 0 : B^r = B^{(r)}$.*

*Proof.* Given $e'$ executed by $SP$, we show that an execution of $P$ that always decides $B^r = B^{(r)}$ is legal. If we can find some legal $e'$ where $\textit{Verify}(B^{(r)}, C^{(r)}) = \textit{False}$ for some $r$, then the inductively constructed $e$ satisfies $\textit{Verify}(B^r, C^r) = \textit{False}$, deeming it illegal. By showing this can never be the case, we prove $e$ is legal and $SP$ is safe.

We have shown in Lemma 3.4.1 that $B^0 = B^{(0)}$ and the genesis block is assumed to be valid. Let us assume, for the sake of contradiction, that an execution $e'$ of $SP$ in which there is some round $r > 0$ where $\textit{Verify}(B^{(r)}, C^{(r)}) = \textit{False}$ exists. If several such rounds exist, let $r$ denote the earliest one $\Rightarrow \forall r' \leq r-1 : \textit{Verify}(B^{(r')}, C^{(r')}) = \textit{True}$ holds. Therefore, we can inductively construct $e$, an execution of $P$ that reaches round $r$ with $C^r = B^0 \circ \ldots \circ B^{r-1} = B^{(0)} \circ \ldots \circ B^{(r-1)} = C^{(r)}$ and $\forall r' \leq r-1 : \textit{Verify}(B^{r'}, C^{r'}) = \textit{True}$.

**Claim 3.4.3.** $\forall i \in [m] : \textit{Verify}(B_i^r, C^{(r)}) = \textit{True}$

*Proof of claim 3.4.3 :* $\textit{Verify}(B_i^r, C_i^r) = \textit{True}$ must hold, because $B_i^r$ is the output of an instance of $P$ running with *honest majority* under context $C_i^r$. Additionally, *self containment* holds, ergo $\textit{Supp}(B_i^r, C^{(r)}) \subseteq C_i^r \Rightarrow \textit{Verify}(B_r^i, C_i^r) = \textit{Verify}(B_r^i, C^{(r)}) = \textit{True}$. □

Back to the main proof, by assumption $\textit{Verify}(B^{(r)}, C^{(r)}) = \textit{False}$. Therefore, the distributed block $B^{(r)}$ makes a conflicting set. Let $X \subseteq B^{(r)}$ be a minimal conflicting set. From Claim 3.4.3 it must be that $X$ contains items from at least two distinct blocks, $B_i^r, B_j^r$ and $i \neq j$.

Let us pick $tx_i, tx_j \in X$ such that $tx_i \in B_i^r$, $tx_j \in B_j^r$ and $i \neq j$. Since $X$ is *minimal*, $tx_i, tx_j$ are competing transactions and must belong to the same partition, as Part is *conflict preserving*. However, we specifically selected $tx_i \in B_i^r \subseteq TX_i$ and $tx_j \in B_j^r \subseteq TX_j$ where $i \neq j \Rightarrow TX_i \cap TX_j = \emptyset$, as Part is a *partition function*. This is a contradiction. □

**Theorem 3.2** ($SP = P$).
*For any legal execution of $P$ there exists a legal equivalent execution of $SP$ and vice versa.*

*Proof.* Lemma 3.4.1 above gives $P \subseteq SP$ and Lemma 3.4.2 gives $SP \subseteq P$. □

## Chapter 4

# Applying the Framework

In this chapter the framework is put to its first use. We demonstrate the steps required to properly apply the framework, create a sharded protocol and to assert correctness:

First, an adequate protocol to be sharded is selected. Next, implementations of all required interfaces are specified and their correctness is proved. Finally, a specific value for $h$ is set and the values $m$ can take are analyzed. By following these steps, the sharded protocol is well defined and its correctness is fully asserted.

The selected blockchain protocol for our demonstration is Algorand.

## 4.1 Related Work - Algorand

### 4.1.1 Algorand in a Nutshell

Algorand, by Micali et al., [7, 14] is an efficient PoS blockchain protocol with negligible probability of creating forks. Its efficiency stems mostly from blocks being admitted to the chain by means of cryptographic sortition, rather than a network-wide race to find PoW. Adjusted for their stake in the system, users are randomly sampled either as block proposers or as committee members, once per round. Only a handful of users may be selected as block proposers. Each selection comes with a priority and the highest priority proposer is the round leader. The expected committee size is larger and committee members run a Byzantine Agreement protocol named $BA\star$[5] to agree on the leader's identity, the proposed block and its admissibility. If any fails, consensus on the empty block is reached.

Algorand's cryptographic sortition mechanism is implemented by verifiable random functions (VRFs)[15]. A selection procedure, executed locally and non-interactively by the user itself, returns a result $res$ and a proof for that result $\pi_u^r$. Selected users may claim their role as block proposers or committee members by revealing $\langle res, \pi_u^r \rangle$ as credentials. Specifically: $\langle res, \pi_u^r \rangle \leftarrow \texttt{Sortition}((PK_u, SK_u), seed^r, role)$ requires a global randomness seed along with the user's cryptographic keys in order to create $\pi_u^r$, a proof tying a randomly sampled $res$ with $PK_u$, $role$ and round $r$. The authenticity of any claimed $\langle res, \pi_u^r \rangle$ pair may be verified by $\texttt{VerifySort}(PK_u, seed^r, role, res, \pi_u^r)$.

Note that $\langle res, \pi_u^r \rangle$ is publicly verifiable, anyone can execute `VerifySort`, as it takes only public parameters. Since `Sortition` is non-interactive and takes privately kept data, known only to the user ($SK_u$), selection is *discrete*. This helps to mitigate targeted attacks. Algorand's sortition is considered tamper-proof since a $\langle res, \pi_u^r \rangle$ pair generated in any way other than invoking `Sortition`$((PK_u, SK_u), seed^r, role)$ properly, will not pass `VerifySort`$(PK_u, seed^r, role, res, \pi_u^r)$ with non-negligible probability.

### 4.1.2 Algorand's VRF Construction

The following components are assumed to be globally known and publicly available:
- $H$ - a cryptographic hash function that doubles as a random oracle.
- `SIG` - a unique digital signature scheme.
- $seed^r$ - a random seed, available at time $r$.

Then: $\pi_u^r = \text{Sign}_{SK_u}((seed^r, role))$ and $res = H(\pi_u^r)$.

More specifically: a user $u$, whose $PK_u \in PK^r$, can generate a unique $\pi_u^r$ by signing the string $(seed^r, role)$ using `SIG`. Then, $res$ is created simply by hashing $\pi_u^r$. Clearly only $u$ can generate a valid $\pi_u^r$, but anyone can verify $\pi_u^r$ by calling $\text{Ver}_{PK_u}((seed^r, role), \pi_u^r)$. Then, $res$ is easily asserted by re-hashing $\pi_u^r$.

$PK_u$ is selected as a block proposer if $res = H(\text{Sign}_{SK_u}(seed^r, proposer)) \leq \tau_{proposer}$ where $\tau_{proposer}$ is a known threshold. The leader is the user with minimal $res$ value. Committee members are similarly selected. Since $H$ is a random oracle and $seed^r$ is unpredictable, the output of each $H(\text{Sign}_{SK_u}(seed^r, role))$ is uniformly distributed and users are randomly sampled.

The initial $seed^0$ is bootstrapped using some distributed random number generation algorithm. The algorithm is collectively carried by the initial set of users, once genesis time users register their public keys. Succeeding rounds' random seeds are efficiently generated by setting either $seed^{r+1} \leftarrow VRF_{SK_u}(seed^r \circ r + 1)$ when $PK_u$ is the round leader and proposed a valid $B^r$ or $seed^{r+1} \leftarrow H(seed^r \circ r + 1)$ if $BA\star$ decided $B^r = \bot$.

### 4.1.3 Algorand's Adversarial Model

Algorand can withstand an adaptive poly-time adversary that controls users holding up to 1/3 of overall funds in the network. This adversary is seen as a single entity with perfect coordination and total control over the actions of all malicious participants. It can initiate an attack and turn an honest participant malicious within a single round ($t_{takeover} = 1$). In order for Algorand to operate properly, at least 2/3 stake in the system must always be under honest ownership, i.e. $h_{Algorand} = 2/3$.

## 4.2 Sharding Algorand - Introducing Algoshard

Two sharded variants of Algorand, *Algoshard0* and *Algoshard1*, are introduced in this chapter. This section provides a high level description and intuition regarding each protocol. Actual implementations appear in section 4.3, in which both versions of Algoshard are presented side by side, followed by proofs of their correctness.

We set $h = 3/4$ in both variants, suited for $h_{Algorand} = 2/3$. The same `Partition` module is reused by both variants, who split the workload between shards identically. Each variant has its own `Sync` and `Membership` implementations, suited for different adversarial assumptions, as detailed below.

### 4.2.1 Algoshard0

Algoshard0 abides Algorand's strict adversarial model, specifically setting $t_{takeover} = 1$. To withstand such a powerful adversary, one able to corrupt any participants within a single round, membership leases are ephemeral. Implemented in `Eager Membership`, every new round sees a new uniform assignment of all participants into all shards. [1] Assignments are randomly drawn, new assignments are independent of previous ones. Since Algoshard0 nodes continuously hop from one random shard to another, never standing still, swift and unpredictable shard transitions must be accommodated. This is achieved by storing the entire global state locally, as implemented in `Eager Sync`.

In that sense, Algoshard0 is logically sharded rather than physically sharded. The block creation process is successfully parallelized, yet the global state remains fully replicated by all nodes. Overall, Algoshard0 can scale throughput whilst resisting an extremely able adversary.

### 4.2.2 Algoshard1

Algoshard1 caters for a more relaxed adversary, with a considerable $t_{takeover}$, spanning an order of one day for example. This setting makes way for longer lasting membership leases, as attacks take longer to complete. Implemented by `Lazy Membership`, the protocol's execution begins with a uniform assignment of all participants into all shards. Then, existing assignments gradually shift, as a small set of participants is required to attain new (uniformly selected) memberships in order to continue participating in the protocol. Memberships are sticky and cannot be manipulated by an adversary. Upon joining the system, each participant is assigned a sort of "personalized membership epoch" timeframe that sprawl specific rounds. Repeated invocations of *getMembership* within the same epoch always return the same result. New users must sit out their first epoch, before being permitted to participate. This helps to mitigate join-leave attacks.

---

[1] To be precise, a different certificate could be set for every step in every round of Algorand. This follows immediately from Algorand's player replaceability property.

By definition, members of $P_i$ need nothing more than the (global) support of $TX_i$ in order to execute $P_i$ correctly. Since `Lazy Membership` puts lengthy memberships in place, swiftness of transition and thrashing are no longer much of a concern. Thus, local storage of the entire global state becomes redundant. This is where `Lazy Sync` comes into play, as it collects only the missing remote support items of $TX_i$, those needed to maintain $Supp(TX_i)$. Then, local states become proper subsets of the global state, for any value of $m > 1$. Moreover, our selection of `Part` suggests that larger values of $m$ reduce the relative size of local states, since each shard stores fewer historical records [2]. Moving shards essentially means re-joining the system, a task that may take a few rounds to complete. Still, this task is infrequent and randomly distributed between the nodes (epochs are uniformly assigned), therefore should not be obstructing operability.

Overall, Algoshard1 is both physically and logically sharded. The block creation process is parallelized while also local states are proper subsets of the global state. Adversarial assumptions are somewhat relaxed in comparison to Algorand, still the adversary is extremely apt.

The analysis of Algoshard's `Membership` implementation, provided in section 4.3.3, shows that networks with more participants support larger values of $m$. In densely populated networks, this implies both increased throughput and reduced local contexts sizes. Assuming that networks with heavier activity and heftier ledgers also contain more participants, Algoshard1 truly scales.

## 4.3 Assembling Algoshard - Implementation and Correctness

*Algoshard0* and *Algoshard1* implementations are introduced side by side.
- `Partition` (section 4.3.1) is shared by both protocols. This section is not divided.
- `Eager Sync` (in 4.3.2) and `Eager Membership` (in 4.3.3) implement Algoshard0.
  – Both appear on the left side of the relevant section.
- `Lazy Sync` (in 4.3.2) and `Lazy Membership` (in 4.3.3) implement Algoshard1.
  – Both appear on the right side of the relevant section.

Proof of each interface correctness follows its implementation.

In section 4.3.3 we analyze how to properly select adequate values for $m$.

*Remark.* Provided implementations (and proofs) for both `Partition` and `Sync` are generic and may be reused as-is with any suitable protocol. However, `Membership`'s implementation is specifically tailored to Algorand, as it piggybacks on the public randomness Algorand already creates. A generic version for this implementation may be created too, for example by incorporating a standalone public randomness mechanism.

---

[2] Assuming that transactions are uniform between all shards' clients, each local state is roughly $2/m - 1/m^2$ the size of the global state.

### 4.3.1 Partition

For ease of exposition we transform *PKS*, the public-key space, into the interval $(0, 1]$.

---
**Algorithm 4:** `Partition` implementation

// $\forall i \in [m]$ : `define` $PKS_i = (\frac{i-1}{m}, \frac{i}{m}]$

1   $\mathsf{Part}_m(T \subseteq TX) = \mathsf{Part}(T, m) = \{T_1, \ldots, T_m\}$ s.t. $T_i = \{tx \in T \mid tx.from \in PKS_i\}$

2   $whichPart_m(tx) = \lceil (tx.from) \cdot m \rceil$

---

Intuitively, *PKS* is lexicographically split into $m$ disjoint intervals of equal sizes, $PKS_1, \ldots, PKS_m$. Transactions sent *from* users whose public keys reside in $PKS_i$ fall within $TX_i$ and therefore pend $P_i$'s approval.

Going forward, we refer to users whose public keys falls in $PKS_i$ as $P_i$'s clients.

**Proposition 4.3.1.** *Part is a partition function*

*Proof.* According to definition 2.1, any $tx \in TX$ must include exactly one $PK_{from} \in PKS$ under *tx.from*. Take any $PK_u \in PKS$, it must fall within the bounds of exactly one subinterval. This gives $TX = \bigcup_{i \in [m]} TX_i$ and $\forall i \neq j : TX_i \cap TX_j = \emptyset$. Additionally, $\forall i \in [m] : \exists PK_u \in PKS_i \Rightarrow \exists tx \in TX$ s.t. $tx.from = PK_u$ which means $TX_i \neq \emptyset$ and completes the proof. $\square$

**Proposition 4.3.2.** *Part is conflict preserving*

*Proof.* In both the *UTXO* model and the account balance model, an object is used as input to a transaction. This object has a single[3] owner authorized to sign for it, the owner of that account or of that *UTXO* set. All transactions sent by $u$ always fall within the same $TX_i$ partition, where $i = \lceil (PK_u) \cdot m \rceil$. This aggregates all conflicts that may arise when several transactions clash over the same asset. Formally:

Let $X \subseteq TX$ be some set of transactions such that $tx_1, tx_2 \notin X$ and $Verify(X \cup \{tx_1\}, C) = True$, $Verify(X \cup \{tx_2\}, C) = True$ but $Verify(X \cup \{tx_1, tx_2\}, C) = False$. Then, by definition, $tx_1, tx_2$ are competing transactions. Let us prove that $\mathsf{Part}$ assigns both $tx_1, tx_2$ into the same partition, in both schemes we consider possible:

**UTXO scheme:** Since either one of the pair $tx_1, tx_2$ could be admitted, but not both, there must be some $utx \in UTXO$ that doubles as an input to both transactions. The owner of $utx$ is the only user authorized to sign and use it, which implies $tx_1.from = tx_2.from = utx.to = PK_u$. This immediately gives $tx_1, tx_2 \in TX_{\lceil PK_u \cdot m \rceil}$

**Balance scheme:** Similar analysis leads to concluding that $PK_u$ has sufficient funds for either $tx_1$ or $tx_2$ but (when combined with $X$) it cannot afford both $\Rightarrow$ $tx_1, tx_2 \in TX_{\lceil PK_u \cdot m \rceil}$

$\square$

---
[3] Recall, we assume that each transaction has a single sender. We implicitly assume that transactions from multiple senders / public keys are aggregated into a multisig address (requiring *m-of-n* signatures).

### 4.3.2 Sync

*Eager collectSupport* reconstructs a global block locally by collecting missing sub-blocks. *Lazy collectSupport* collects only transactions sent *to* $P_i$'s clients and missing from its local block. Both `Sync` implementations are suitable for `Partition` presented in 4.3.1.

**Algorithm 5:** *Eager collectSupport*
- **Input**: $B^{(r)} \subseteq TX, i \in [m]$
- **Output**: $\{B^{(r)} \setminus B_i^r\} = \{tx \in B^{(r)} \mid tx.from \notin PKS_i\}$
1. $RS_i^r = \emptyset$;
2. **foreach** $tx \in B^{(r)}$ **do**
3.    **if** $tx.from \notin PKS_i$ **then** $RS_i^r = RS_i^r \cup \{tx\}$;
4. **return** $RS_i^r$;

**Algorithm 6:** *Lazy collectSupport*
- **Input**: $B^{(r)} \subseteq TX, i \in [m]$
- **Output**: $\{tx \in B^{(r)} \mid tx.from \notin PKS_i \wedge tx.to \in PKS_i\}$
1. $RS_i^r = \emptyset$;
2. **foreach** $tx \in B^{(r)}$ **do**
3.    **if** $tx.from \notin PKS_i \wedge tx.to \in PKS_i$ **then** $RS_i^r = RS_i^r \cup \{tx\}$;
4. **return** $RS_i^r$;

Algoshard0 uses *Eager collectSupport* as it enables participants to maintain the full global state. This makes swift and unpredictable transitions between shards possible.

**Proposition 4.3.3.** *Eager collectSupport guarantees self containment throughout the entire execution:* $\forall i \in [m], r \in \mathbb{N} : Supp(TX_i, C^{(r)}) \subseteq C_i^r$

*Proof.* By definition, *Eager collectSupport* returns $RS_i^r = \{B^{(r)} \setminus B_i^r\}$, for all $i \in [m]$. It follows by induction that $C_i^r = C^{(r)}$ :

$r = 0 : C_i^0 = \bot = C^{(0)}$

$r > 0 : C_i^r = C_i^{r-1} \circ B_i^{r-1} \circ RS_i^{r-1} = \underbrace{C_i^{r-1}}_{C^{(r-1)}} \circ \underbrace{B_i^{r-1} \circ \{B^{(r-1)} \setminus B_i^{r-1}\}}_{B^{(r-1)}} = C^{(r)}$

$\Rightarrow \forall X \subseteq TX$ and in particular $\forall X \subseteq TX_i : Supp(X, C^{(r)}) \subseteq C^{(r)} = C_i^r$ . $\square$

Algoshard1 uses *Lazy collectSupport,* a construction that returns a more compact set, resulting in lighter local states. The following proof establishes that when local contexts are kept up-to-date by using *Lazy collectSupport*, self containment is preserved.

**Proposition 4.3.4.** *Lazy collectSupport guarantees that self containment is kept.*
It holds that: $RemoteSupp_i(B^{(r)}) \subseteq \{tx \in B^{(r)} \mid tx.from \notin PKS_i \wedge tx.to \in PKS_i\} = RS_i^r$.

*Proof.* We establish $Supp(TX_i, C^r) \subseteq \{tx \in TX \mid tx.from \in PKS_i \vee tx.to \in PKS_i\} \cap C^r$ first. Then, we show $RemoteSupp_i(B^{(r)}) \subseteq \{tx \in B^{(r)} \mid tx.from \notin PKS_i \wedge tx.to \in PKS_i\}$.

**Claim 4.3.5.** *for any $X \subseteq TX$ and any legal $C^r$* :
$Supp(X, C^r) \subseteq \{tx \in C^r \mid \exists t \in X \ s.t. \ tx.from = t.from \vee tx.to = t.from\}$

*Proof of claim 4.3.5 :*

Define $Supp'(X) = \{tx \in TX \mid \exists t \in X \text{ s.t. } tx.from = t.from \lor tx.to = t.from\}$.
It suffices to show that $Verify(X, C^r) = Verify(X, Supp'(X) \cap C^r)$ always holds.

Let $L_u = \{tx \in TX \mid tx.from = PK_u \lor tx.to = PK_u\}$. Then, $HIST_u(C^r) \triangleq L_u \cap C^r$ is the set of all approved transactions sent by or sent to $PK_u$ up to time $r$ [4]. Any bookkeeping method used to determine either $PK_u$'s balance or set of $UTXO$ must return identical results with either $C^r$ or $HIST_u(C^r)$ as input. Now, let us expand $L_u$'s definition to sets of users by $L_U \triangleq \bigcup_{u \in U} L_u$.

For any $X \subseteq TX$ let us define $PK_X = \{PK_u \in PKS \mid \exists t \in X \text{ s.t. } t.from = PK_u\}$, the set of users that *send* payments in $X$. Then, $HIST_{PK_X}(C^r) = L_{PK_X} \cap C^r$ is the complete log of all sent and received transactions by members of $PK_X$ since genesis. This log suffices to deduce the complete set of their up-to-date resources, as appearing in $C^r$. Therefore $X$, as well as any other set of transactions sent from $PK_X$'s members, satisfies $Verify(X, C^r) = Verify(X, HIST_{PK_X}(C^r))$. Note that $HIST_{PK_X}(C^r) = Supp'(X) \cap C^r$.
Overall, $Verify(X, C^r) = Verify(X, Supp'(X) \cap C^r)$ holds. $\square$

**Corollary 4.1.** $Supp(TX_i, C^r) \subseteq \{tx \in TX \mid tx.from \in PKS_i \lor tx.to \in PKS_i\} \cap C^r$.
*It is a direct result of claim 4.3.5, simply by plugging in $X = TX_i$.*

Let us define: $LSupp(TX_i) = \{tx \in TX \mid tx.from \in PKS_i\}$
$\qquad\qquad\quad RSupp(TX_i) = \{tx \in TX \mid tx.from \notin PKS_i \land tx.to \in PKS_i\}$

$LSupp(TX_i) \cap RSupp(TX_i) = \emptyset$ and
$LSupp(TX_i) \cup RSupp(TX_i) = \{tx \in TX \mid tx.from \in PKS_i \lor tx.to \in PKS_i\} = Supp'(TX_i)$
both hold. Observe that $B^{(r)} \cap LSupp(TX_i) = B_i^r$, by 4.3.1. Corollary 4.1 gives:
$Supp'(TX_i) \cap B^{(r)} \subseteq \{tx \in B^{(r)} \mid tx.from \in PKS_i \lor tx.to \in PKS_i\}$.
These give way to $RemoteSupp_i(B^{(r)}) = Supp(TX_i) \cap \{B^{(r)} \setminus B_i^r\} \subseteq RSupp(TX_i) \cap B^{(r)} = \{tx \in B^{(r)} \mid tx.from \notin PKS_i \land tx.to \in PKS_i\}$ which completes the main proof. $\square$

---

[4] Generally speaking, this makes the set of all resources attained and released by $PK_u$ up to time $r$.

### 4.3.3 Membership

A key part of Algoshard's construction, as well as the reason for surveying Algorand's VRFs in-depth earlier, is implementing Membership by a VRF construction very similar to Algorand's sortition procedure.

The general idea behind Algoshard's Membership is the following:
A global random seed, common to all participants and all shards, is maintained throughout the execution. Users locally generate their own shard assignments and membership certificates. Users propagate their certificates, for others to verify, whenever necessary. A certificate $\sigma_u$ is created by computing $\sigma_u = \text{Sign}_{SK_u}(seed)$. Then, a uniform shard is selected by using $\sigma_u$ as an input to $.H$, thereby binding $u$ with $P_i$ where $i = \lceil .H(\sigma_u) \cdot m \rceil$. Assuming that $seed$ is indeed random, SIG is a unique signature scheme and $.H$ is a cryptographic hash function implies that $.H(\sigma_u)$ yields a uniformly drawn (binary) value between 0 and 1. This, in turn, implies that $i \in [m]$ is indeed randomly drawn. Any user-claimed assignment is easily verifiable, simply by executing $\text{Ver}_{PK_u}(seed, \sigma_u)$ and recalculating $\lceil .H(\sigma_u) \cdot m \rceil$.

The binding between participants and assigned shards is time-restricted, thereby averting targeted attacks on specific shard(s). Restriction details are variant dependant: In EagerMembership, used by Algoshard0, certificates are valid for a single round. [5]
LazyMembership, used by Algoshard1, begins by bootstrapping a uniform assignment of all initial participants into all shards. It then proceeds by slowly shuffling current assignments. A global parameter $1 \leq t_{lease} \leq t_{takeover}$ is set by the deployer. Upon joining the network, $PK_u$ is publicly and randomly assigned a slot $t_u^{shuffle} \in [0, t_{lease}-1]$. Every $t_{lease}$ rounds, starting from $t_u^{shuffle}$, its previous certificate expires and $PK_u$ must acquire a new membership lease in order to continue participating in the protocol. New participants must wait $t_{lease}$ rounds before their $t_u^{shuffle}$ slot is set and their first shard assignment is known and comes into effect. This resolves join-leave attacks.

Throughout Algoshard's execution, participants essentially invoke one VRF in order to select which second VRF they should call. $\langle i_u, \sigma_u \rangle \leftarrow getMembership((PK_u, SK_u), r)$ followed by Algorand's $\langle res, \pi_u^r \rangle \leftarrow \text{Sortition}((PK_u, SK_u), seed_{i_u}^r, role)$ are both needed. That is because $P_i$'s members simply discard $u$'s actions, even those sent with $\langle res, \pi_u^r \rangle$ that allows $u$ to act as block proposer or verifier, unless $\sigma_u$ that permits $u$ to participate in $P_i$ is also provided.

---

[5]EagerMembership is actually a special, simpler, case of LazyMembership where $t_{lease} = 1$. EagerMembership's algorithms are explicitly included, as it might be easier to follow this much simpler variant before proceeding to the general version.

**Bootstrapping** `Membership`

| **Algorithm 7:** EagerMembership init | **Algorithm 8:** LazyMembership init |
|---|---|
| **Input** : $m \in \mathbb{N}^+, PK^0 \subseteq PKS$ <br> **Output** : participants bookkeeping | **Input** : $m \in \mathbb{N}^+, PK^0 \subseteq PKS, t_{lease} \in \mathbb{N}^+$ <br> **Output** : participants bookkeeping |
| 1  $seed^0 = AlgorandInitRandomness(PK^0)$; <br> 2  **foreach** $i \in [m]$ **do** <br>     // init $P_i$'s public randomness <br> 3    $seed_i^1 \leftarrow H(seed^0 + i)$ <br> 4  **end** <br> 5  $seed^1 = H(seed_1^1 \circ \ldots \circ seed_m^1)$; <br> 6  **foreach** $PK_u \in PK^0$ **do** <br>     // publicly logged by all nodes <br> 7    $t_{PK_u}^{join} = 0$; <br>     // privately invoked by each participant <br> 8    $\langle i_u^1, \sigma_u^1 \rangle = getMembership((PK_u, SK_u), 1)$; <br> 9    join $P_{i_u^1}$; <br> 10  **end** <br> 11  **return** $\{(PK_u, t_{PK_u}^{join}) \mid PK_u \in PK^0\}$; | 1  $seed^0 = AlgorandInitRandomness(PK^0)$; <br> 2  **foreach** $i \in [m]$ **do** <br>     // init $P_i$'s public randomness <br> 3    $seed_i^1 \leftarrow H(seed^0 + i)$ <br> 4  **end** <br> 5  $seed^1 = H(seed_1^1 \circ \ldots \circ seed_m^1)$; <br> 6  **foreach** $PK_u \in PK^0$ **do** <br>     // publicly logged by all nodes <br> 7    $t_{PK_u}^{join} = 0$; <br> 8    $t_{PK_u}^{shuffle} = H(PK_u \circ seed^1) \pmod{t_{lease}}$; <br>     // privately invoked by each participant <br> 9    $\langle i_u^1, \sigma_u^1 \rangle = getMembership((PK_u, SK_u), 1)$; <br> 10    join $P_{i_u^1}$; <br> 11  **end** <br> 12  **return** $\{(PK_u, t_{PK_u}^{join}, t_{PK_u}^{shuffle}) \mid PK_u \in PK^0\}$; |

Once genesis time participants register their public keys, one initial random seed is selected. This $seed^0$ is generated by genesis time users calling the same distributed random numbers generation algorithm Algorand employs. $seed^0$ is then used to kick-start $m$ independent instances of Algorand, rather than only one. This is efficiently done by setting $P_i$'s initial public randomness to be $seed_i^1 = H(seed^0 + i)$, which is indeed random since $seed^0$ is randomly selected and $H$ is a random oracle.

Both `Membership` variants begin with a uniform assignment of all participants into all shards. The initial assignment depends on $seed^1$, whose value is unknown at the time participants register their public keys. In `LazyMembership`, each $PK_u$ is publicly and randomly assigned a permanent $t_u^{shuffle} \in [0, t_{lease} - 1]$ slot. Note that $t_u^{shuffle}$ is publicly verifiable and it is always determined by a random seed that is still unknown when $u$ registers $PK_u$.

The following $m + 1$ public randomness seeds are maintained throughout the execution:

- $seed_1^r, \ldots, seed_m^r$ - local public randomness seeds.
  $seed_i^r$ is the public randomness used by $P_i$. Its progression follows Algorand's standard definition: $seed_i^{r+1}$ is either $VRF_{SK_u}(seed_i^r \circ r + 1)$ or $H(seed_i^r \circ r + 1)$, depending on $B_i^r$.

- $seed^r$ - a global public randomness seed, serves as input to `Membership`.
  $seed^r \triangleq H(seed_1^r \circ \ldots \circ seed_m^r)$ is created by bundling all $m$ local public seeds.

## VRF construction

**Algorithm 9:** *Eager getMembership*

    **Input**    : $(PK_u, SK_u), r$
    **Output** : $\langle i_u^r, \sigma_u^r \rangle$

1. $\sigma_u^r = \text{Sign}(SK_u, seed^r)$;
2. $h_u^r = .H(\sigma_u^r)$;
3. $i_u^r = \lceil h_u^r \cdot m \rceil$;
4. **return** $\langle i_u^r, \sigma_u^r \rangle$;

**Algorithm 10:** *Lazy getMembership*

    **Input**    : $(PK_u, SK_u), r$
    **Output** : $\langle i_u^r, \sigma_u^r \rangle$

1. $slot = r \% t_{lease}$;
2. $diff = slot - t_u^{shuffle} \ (mod \ t_{lease})$;
3. $r' = r - diff$;
4. $\sigma_u^r = \text{Sign}(SK_u, seed^{r'})$;
5. $h_u^r = .H(\sigma_u^r)$;
6. $i_u^r = \lceil h_u^r \cdot m \rceil$;
7. **return** $\langle i_u^r, \sigma_u^r \rangle$;

To participate in Algoshard, $u$ must acquire membership in some $P_i$ and a proof thereof. This is done by executing $getMembership((PK_u, SK_u), r) \to \langle i_u^r, \sigma_u^r \rangle$ locally and non-interactively. Since $seed^r$ is unpredictable and $H$ is a random oracle, $h_u^r = .H(\sigma_u^r)$ is uniform between 0 and 1 which makes $i_u^r \in [m]$ a uniformly selected shard index.

**Algorithm 11:** *Eager verifyMember*

    **Input**    : $PK_u, \sigma_u^r, i, r$
    **Output** : *True*   $\sigma_u^r$ proves $PK_u$'s $P_i$
                    membership in round $r$
                *False*  otherwise

1. **if** $\lceil .H(\sigma_u^r) \cdot m \rceil \neq i$ **then return** *False*;
2. **if** $PK_u \notin PK^{r-1}$ **then return** *False*;
3. **return** $\text{Ver}(PK_u, seed^r, \sigma_u^r)$;

**Algorithm 12:** *Lazy verifyMember*

    **Input**    : $PK_u, \sigma_u^r, i, r$
    **Output** : *True*   $\sigma_u^r$ proves $PK_u$'s $P_i$
                    membership in round $r$
                *False*  otherwise

1. **if** $\lceil .H(\sigma_u) \cdot m \rceil \neq i$ **then return** *False*;
2. **if** $PK_u \notin PK^{max\{0, r-t_{lease}\}}$ **then return** *False*;
3. attain $t_{PK_u}^{shuffle}$;
4. $slot = r \% t_{lease}$;
5. $diff = slot - t_{PK_u}^{shuffle} \ (mod \ t_{lease})$;
6. $r' = r - diff$;
7. **return** $\text{Ver}(PK_u, seed^{r'}, \sigma_u)$;

By adopting Algorand's cryptographic assumptions regarding $\text{SIG}$ and $H$, it follows that $\langle i_u^r, \sigma_u^r \rangle$ is unforgeable by a poly-time adversary.

It remains to prove the resulting assignments are indeed safe, even when facing an adaptive adversary. This is done in proposition 4.3.6.

## Additional functionalities

**Algorithm 13:** *Eager newNodesArrival*

    **Input** : $r \in \mathbb{N}^+, PK_{new}^r \subseteq PKS$
    **Output** : new participants bookkeeping

    // publicly logged by all nodes
1 **foreach** $PK_u \in PK_{new}^r$ **do**
2     $t_{PK_u}^{join} = r$;
3 **end**
4 **return** $\{(PK_u, t_{PK_u}^{join}) \mid PK_u \in PK_{new}^r\}$;

---

**Algorithm 14:** *Lazy newNodesArrival*

    **Input** : $r \in \mathbb{N}^+, PK_{new}^r \subseteq PKS, t_{lease} \in \mathbb{N}^+$
    **Output** : new participants bookkeeping

    // publicly logged by all nodes
1 **foreach** $PK_u \in PK_{new}^r$ **do**
    // recorded at time $r$
2     $t_{PK_u}^{join} = r$;
    // set after benching $PK_u$ $t_{lease}$ rounds
3     $t_{PK_u}^{shuffle} = H(PK_u \circ seed^{t_{PK_u}^{join}+t_{lease}}) \ (mod \ t_{lease})$;
4 **end**
5 **return** $\{(PK_u, t_{PK_u}^{join}, t_{PK_u}^{shuffle}) \mid PK_u \in PK_{new}^r\}$;

---

**Algorithm 15:** *Eager endOfRound*

    **Input** : $r, seed_1^{r+1}, \ldots, seed_m^{r+1}$
    **Output** : $seed^{r+1}$, updated membership

    // locally executed by each node
1 $seed^{r+1} = H(seed_1^{r+1} \circ \ldots \circ seed_m^{r+1})$;
2 $\langle i_u^{r+1}, \sigma_u^{r+1} \rangle =$
    $getMembership((PK_u, SK_u), r+1)$;
3 leave $P_{i_u^r}$ and join $P_{i_u^{r+1}}$ ;
4 **return** $seed^{r+1}$;

---

**Algorithm 16:** *Lazy endOfRound*

    **Input** : $r, seed_1^{r+1}, \ldots, seed_m^{r+1}$
    **Output** : $seed^{r+1}$, updated membership

    // locally executed by each node
1 $seed^{r+1} = H(seed_1^{r+1} \circ \ldots \circ seed_m^{r+1})$;
2 **if** $t_{PK_u}^{shuffle} \equiv r+1 \ (mod \ t_{lease})$ **then**
3     $\langle i_u^{r+1}, \sigma_u^{r+1} \rangle =$
    $getMembership((PK_u, SK_u), r+1)$;
4     leave current $P_i$ and join $P_{i_u^{r+1}}$;
5 **end**
6 **return** $seed^{r+1}$;

**Proposition 4.3.6.** *Given $n$ participants, of which at least $\frac{3n}{4}$ are honest, an appropriately set parameter $m$ and faced with an adaptive adversary that is able to corrupt participants in $t_{takeover}$ rounds: both* `EagerMembership` *and* `LazyMembership` *guarantee an honest majority of $h_{Algorand} = 2/3$ in all shards with extremely high probability.*

**Membership Correctness**

**Transitioning into balls and bins** Consider representing an honest participant by a blue ball, a Byzantine participant by a red ball and each $P_i$ by a bin. Then the analysis of failure could be seen as a model with $\frac{3n}{4}$ blue balls and $\frac{n}{4}$ red balls placed into $m$ bins, according to Membership's logic. If a state in which the number of blue balls in some bin is less than twice the number of red balls in that bin is reached, we lose, *honest majority* breaks. We bound this probability throughout the execution. Let us begin:

The initial uniform assignment of both Membership variants goes as follows:
All $n$ balls, regardless of color, are independently and uniformly tossed into all $m$ bins.

**Lemma 4.3.7.** *Let us define failure as any of the bins having a red to overall balls ratio greater than or equal to $1/3$. In a setting with $\frac{3n}{4}$ blue balls and $\frac{n}{4}$ red balls, all uniformly assigned into $m$ bins, it holds that $Pr(failure) < 2me^{-\frac{n}{144m}}$.*

*Proof.* Let us examine the situation in the $i^{th}$ bin (analysis is symmetric for all bins and we union bound it later). From the bin's perspective, any tossed ball is a Bernoulli trial with success probability $p = Pr(X = 1) = \frac{1}{m}$ and all tosses are independent. We denote $Y$ as the number of blue balls assigned to our bin and $Z$ as the number of red balls assigned to our bin. We get that $Y = Y_1 + Y_2 + ... + Y_{\frac{3n}{4}}$ where $Y_j$ represents the $j^{th}$ blue Bernoulli experiment and $Z = Z_1 + Z_2 + ... + Z_{\frac{n}{4}}$ where $Z_j$ represents the $j^{th}$ red Bernoulli experiment, alternatively $Y \sim Bin(\frac{3n}{4}, \frac{1}{m}), Z \sim Bin(\frac{n}{4}, \frac{1}{m})$. In both cases we have that $E[Y] = \frac{3n}{4m}$ and $E[Z] = \frac{n}{4m}$. We define $X = Y + Z$ as the total number of balls in the bin. Clearly $Y, Z$ are independent (as all tosses are i.i.d.) and then $X \sim Bin(n, \frac{1}{m})$ with $E[X] = \frac{n}{m}$. Complying with the law of large numbers, as $n \to \infty$ and while keeping $m$ constant or sufficiently small, $Y \to \frac{3n}{4m}$ and $Z \to \frac{n}{4m}$. The expected ratio of red to overall balls in the bin is $E[\frac{Z}{Y+Z}] = \frac{E[Z]}{E[Y]+E[Z]} = \frac{1}{4}$. We may deviate somewhat from the mean and still reach an acceptable (non-failed) state. In order to estimate how likely it is to deviate too far, we now bound the tail probability $Pr(\frac{Z}{Y+Z} \geq \frac{1}{3})$ of states in which the system breaks, as we cannot safely run $P_i$ with too many Byzantine players.

1. $Pr(Y \leq \dfrac{2.5n}{4m}) \leq e^{-\frac{n}{96m}}$

    - The Chernoff bound $Pr(X \leq (1-\delta)\mu) \leq e^{-\frac{\delta^2 \mu}{2}}$ applied on $Y$ with $\delta = \frac{1}{6}$ and $\mu = E[Y] = \frac{3n}{4m}$ gives
    $Pr(Y \leq \frac{2.5n}{4m}) = Pr(Y \leq (1 - \frac{1}{6})\frac{3n}{4m}) \leq e^{-\frac{(\frac{1}{6})^2 \cdot \frac{3n}{4m}}{2}} = e^{-\frac{n}{96m}}$.

2. $Pr(Z \geq \dfrac{1.25n}{4m}) \leq e^{-\frac{n}{144m}}$

    - Similarly, using the bound $Pr(X \geq (1+\delta)\mu) \leq e^{-\frac{\delta^2 \mu}{2+\delta}}$ on $Z$ with $\delta = \frac{1}{4}$ and $\mu = E[Z] = \frac{n}{4m}$ gives $Pr(Z \geq \frac{1.25}{4m}) \leq e^{-\frac{(\frac{1}{4})^2 \cdot \frac{n}{4m}}{2+\frac{1}{4}}} = e^{-\frac{n}{144m}}$.

3. $Pr(\frac{Z}{Y+Z} \geq \frac{1}{3}) < 2e^{-\frac{n}{144m}}$.

- Following (1) and (2), the probability of failure from the $i^{th}$ bin perspective satisfies:

$Pr(\frac{Z}{Y+Z} \geq \frac{1}{3}) \overset{(*)}{\leq} 1 - (1-e^{-\frac{n}{96m}})(1-e^{-\frac{n}{144m}}) = e^{-\frac{n}{96m}} + e^{-\frac{n}{144m}} - e^{-\frac{n}{96m}} \cdot e^{-\frac{n}{144m}} \leq e^{-\frac{n}{96m}} + e^{-\frac{n}{144m}} < 2e^{-\frac{n}{144m}}$.

(*) Let us define $A = \{$all assignments where $Y \leq \frac{2.5n}{4m}\}$ and $B = \{$all assignments where $Z \geq \frac{1.25n}{4m}\}$. Then, $\overline{A} = \{$all assignments with $Y > \frac{2.5n}{4m}\}$ is a set of states where there aren't too few blue balls and $\overline{B} = \{$all assignments with $Z < \frac{1.25n}{4m}\}$ is a set of states where there aren't too many red balls. It follows that all assignments in $C \triangleq \overline{A} \cap \overline{B}$ satisfy $\frac{Z}{Y} < \frac{\frac{1.25n}{4m}}{\frac{2.5n}{4m}} = \frac{1}{2} \Rightarrow \frac{Z}{Y+Z} < \frac{1}{3}$ are non-failed states where the ratio between red and blue balls is acceptable. Since $\overline{A}, \overline{B}$ are independent and must both occur in order to land $C$, we encounter $C$ with probability $(1 - Pr(A)) \cdot (1 - Pr(B))$ which is at least $(1-e^{-\frac{n}{96m}})(1-e^{-\frac{n}{144m}})$ according to the bounds in (1) and (2). We've shown that all the states in $C$ are acceptable, implying that all unacceptable (failed) states necessarily reside within its complement, $\overline{C}$ which is bounded by $1 - \underbrace{\underbrace{(1-e^{-\frac{n}{96m}})}_{\overline{A}} \cdot \underbrace{(1-e^{-\frac{n}{144m}})}_{\overline{B}}}_{\overline{C}}$.

4. All bins follow the same distribution, allowing us to union bound (3) over all $m$ bins. We conclude that $Pr(failure) < m \cdot (2e^{-\frac{n}{144m}}) = 2me^{-\frac{n}{144m}}$.

□

**Corollary 4.2.** *With $m$ set appropriately, `EagerMembership` fails only with negligible probability, even when continuously running for over a million years (∗).*

(∗) *for any values of $m \leq 10000 \wedge n \geq 15000m$ we get astronomically small bounds. That is, we can safely support up to 10,000 shards when expecting at least 15,000 participants per shard on average. Similarly, any $m \leq 700$ and $n/m \geq 10000$ adequately bounds smaller networks.*

*Proof.* Running with `EagerMembership`, every new round begins with a new uniform assignment adhering to the analyzed distribution. Plugging in the above numbers gives $Pr(failed\ round \mid m \leq 10000 \wedge n \geq 15000m) < 10^{-40}$. We can also attain $Pr(failed\ round \mid n \geq 10000m) < 2me^{\frac{625}{9}} < 10^{-29} \cdot m$, which bounds the second case with $Pr(failed\ round \mid m \leq 700 \wedge n \geq 10000m) < 10^{-27}$.

Following Algorand's one round per minute rate gives a total of $5.26 \cdot 10^{11}$ rounds when running for one million years. By union bounding any of the above bounds over the course of $5.26 \cdot 10^{11}$ rounds, we are still at most $10^{-15}$ likely to fail. □

**Iterated balls into bins** `LazyMembership` can be viewed as an iterated balls into bins process. When a ball joins this process, it is imprinted with a uniformly drawn number between 0 and $t_{lease} - 1$. During the first iteration, where $r = 1$, all balls are uniformly thrown into $m$ bins. Then, for any iteration $r > 1$, we take all balls whose imprinted numbers equal $r$ modulo $t_{lease}$ out of their bins and re-throw them at random.

**Lemma 4.3.8.** *When dealing with a static adversary ($t_{takeover} = \infty$, balls cannot change color), the balls-into-bins distributing for any round $r > 1$ equals the initial uniform assignment distribution that occurs for $r = 1$.*

*Proof.* The initial distribution encountered at $r = 1$ consists of uniformly assigning balls into bins, in any order. The actual order in which balls are placed in their assigned bins makes no difference. Then, two equivalent ways to reach the state at $r = 2$ are:

1. Uniformly assign all balls into bins, in any order. Remove all of the balls whose imprinted numbers equals 2 from their bins. Reassign removed balls uniformly.

2. Uniformly assign all balls whose imprinted numbers do not equal 2 into bins. Wait. Uniformly assign all remaining balls.

Clearly, option 2 is a uniform assignment of all balls into all bins. Then, the first two iterations are identically distributed, as succeeding iterations are, by induction. □

**Corollary 4.3.** *Faced with a static adversary that can introduce malicious users into the system but cannot attack honest users, `LazyMembership` is as safe as `EagerMembership`.*

**Lemma 4.3.9.** *For any selection of $t_{lease}$ where $1 \leq t_{lease} \leq t_{takeover}$, an adaptive adversary (with a finite $t_{takeover}$) cannot do any better than a static adversary.*

*Proof.* By contradiction. Let us assume, for the sake of contradiction, that some adaptive adversary can plan an attack on a target set $B = \{b_1, ..., b_k\}$ of $k$ blue balls and do better than simply sticking with the same set of red balls it already owns.
We assume that the adversary is at capacity and owns $n/4$ red balls (if this isn't the case, we paint red some randomly selected blue balls and hand those over). Let $R = \{r_1, ..., r_k\}$ be the subset of red balls that must released in order to attack $B$, as selected by the adversary's strategy. When the attack completes, $t_{takeover}$ rounds after it was set in motion, all balls in $B$ are painted red and all balls in $R$ are painted blue. In the meantime, since $t_{lease} \leq t_{takeover}$ all balls in the system, and in particular those in $B$ and $R$, are uniformly reassigned. Recall that once an attack has been set in motion it cannot be aborted and must carry on until completion.

Take any set $B$ the adversary may possibly choose to target given any (poly-time) algorithm. By the time the attack completes, any $b_i \in B$ the adversary gained control over (painted red) is exactly $1/m$ likely to be in any specific bin. At the same time, any $r_i \in R$ the adversary released (painted blue) is also $1/m$ likely to be in that same bin, as all assignments are unpredictable. Contradiction to the assumption that any such dynamic strategy beats a static strategy. □

**Corollary 4.4.** *Faced with either a static or an adaptive adversary,* `LazyMembership` *is as safe as* `EagerMembership`. *Its failure rate adheres the same upper bounds.*

**Conclusion:** This completes the proof of proposition 4.3.6. Correctness of both `Membership` implementations has been asserted, when setting $m \leq \frac{n}{10000}$.

Take note that the bounds we used in lemma 4.3.7 are rather loose.
In practice, simulations with $n/m \approx 1500$ passed hundreds of millions of rounds without failing once.

## Chapter 5

# Additional Related Work

Several recent efforts focus on sharding permissionless blockchain protocols. Common, rather intuitive, guiding lines seem to unify all of these efforts (this thesis included): nodes are scattered between distinct groups, members of each group run some internal protocol among themselves and different groups can communicate with each other.

Nevertheless, it appears that all existing work has taken an ad-hoc approach, by introducing specific sharded protocols, rather than providing a generalized methodology for sharding. When introducing concrete protocols with no previous groundwork, each protocol's correctness must be fully asserted from scratch, a tedious and error-prone task. A modular framework design, like the one introduced in this thesis, seems like an overall preferable approach. It provides additional advantages such as applicability to already-existing protocols, reusable modules that could shard several different protocols without work duplication etc.

## 5.1 Recent Sharded Protocols

### 5.1.1 OmniLedger

OmniLedger [10] is a sharded variant of ByzCoin [9] that follows the aforementioned guidelines. OmniLedger is a UTXO based blockchain protocol. It uses an Algorand-like VRF to elect each epoch's leader, who in turn leads an execution of RandHound [18]. RandHound's output is used to determine a global nodes-into-shards assignment permutation. Transitioning between epochs, swap-out policies are locally set by each shard. OmniLedger introduces a several-rounds-long cross shard transaction approval process. This process is driven by the client. After initializing a cross shard transaction, the user has to gather proofs-of-acceptance for all UTXO inputs, from their respective shards, in order to admit the cross shard transaction. Aborting a cross shard transaction can be done by collecting a proof-of-rejection for one of the UTXO inputs. Presenting a proof-of-rejection for any such input enables to reclaim the funds from another already accepted input. In cases where some input is neither accepted nor rejected, the user must wait. An already-admitted input is marked as spent, it cannot be reclaimed or reused until some proof-of-rejection is provided. Such lock-unlock mechanism is, in essence, very similar to using a multisig address as a payer, after gathering desired funds into that account. In fact, several constructions presented in Algoshard are rather similar to those appearing in OmniLedger. Still, it should be noted that OmniLedger is far less formal and proofs for many of their claims are lacking or missing.

### 5.1.2 RapidChain

RapidChain [19] is a new blockchain protocol that comes sharded by design, rather than an adaptation of some pre-existing protocol. It executes in epochs, within a strongly synchronous environment and assumes a limited slowly adaptive adversary. RapidChain's adversary is rather weak, it can commit only a small constant number of join-leave attacks per epoch. Equivalently, the adversary can attack a constant number of nodes and only during the epoch's end.

RapidChain maintains $k + 1$ committees. The first committee is called a reference committee. The reference committee is in charge of establishing participants identity, generating random strings and setting shard assignments. The other $k$ committees are simply known as shards. They run synchronous consensus protocols among shard members to approve pending transactions. Transaction approval as well the task of storing the entire UTXO set are partitioned between the different shards, according to transactions' ids. RapidChain's approach to cross shard transactions, in which some inputs reside within different partitions, is the following: Once a cross shard $tx$ is submitted, the committee in charge of approving $tx$ creates a new transaction for any UTXO input that resides within another shard. These intermediate transactions are meant for transferring funds from their current shard into the same shard as $tx$. Once all intermediate transactions have been approved, funds are locally available and $tx$ too should be approved. RapidChain's committee-to-committee routing mechanism is patterned after distributed hash table (DHT) design, as in Kademlia [13]. In order to participate in the protocol, participants need to solve PoW puzzles, to "activate" their public keys, this is done once per epoch. It is assumed that computing power held by all participants in the system is equal.

## 5.2 A Few Notes Regarding Cross Shard Transactions

Compared with the approach presented in this thesis, it seems as if both OmniLedger and RapidChain overcomplicate the treatment of cross shard transactions. We manage to avoid the issue altogether, while also allowing all participants to send and receive money from each other. This was accomplished by providing general definitions, fine-tuning the partitions and ultimately maintaining a *double-entry ledger*. Double-entry bookkeeping [3, 4] is a well known bookkeeping system in which every transaction is recorded (at least) twice, once as credit and once as debit [1]. Then, our two-phased round structure boils down to updating the debit column followed by updating the credit column. Since our double-entry ledger is always kept up-to-date and partitioned appropriately, we never partially commit transactions that might need to be aborted, as in OmniLedger [2]. Nor do we need to explicitly initiate complex multi-step transactions with intermediate transfers that must first be completed to determine whether the enclosing transaction should be approved or denied, as in RapidChain. Note that the design presented in RapidChain does not reduce the storage overhead compared with our solution. That is because duplicate transfer transactions are created and locally stored by the receiving shard. Moreover, since the enclosing transaction could fail after some intermediate transfers were completed, new transactions that use these already transferred funds may originate from different shards. In this case, additional intermediate transactions must be created, which means that the storage overhead is actually worse.

---

[1] Equivalently, once as an incoming transaction and once as an outgoing transaction

[2] Transferring money to a multisig address is external to this issue. However, this scenario is fairly easily solvable, for example by using a script requiring *1-of-n* signature to send funds back if a multisig transaction was not admitted within a certain timeframe.

# Chapter 6

# Conclusion

This thesis introduced a general methodology for sharding blockchain protocols.
In chapter 2, a formal model expressing the behavior of blockchain protocols was laid out. In chapter 3, building on top of these definitions, a rigorous sharding scheme was introduced. The scheme is specified while black-boxing some core functionality, making it modular and generic. A proof binding the definitions described in chapter 2 and requirements detailed in chapter 3 asserts the framework's overall correctness. Chapter 4 demonstrates that the requirements details in chapter 3 are indeed feasible, by introducing concrete implementations and creating a sharded variant of Algorand.

## 6.1 Discussion

### 6.1.1 Relaxing Some Assumptions

In chapters 2 and 3 we made a few assumptions regarding an arbitrary blockchain protocol, denoted $P$. These assumptions were made in order to attain stronger results and simplify the algorithms. Let us examine the consequences of removing them:

- **Intra-Block Transactions:** It must be assumed that $P$ does not permit intra-block transactions to obtain $P = SP$.
  Removing this assumption, the $SP \subseteq P$ direction of the proof in section 3.4 still holds, for protocols that do allow intra-block transactions too. This makes such protocols' sharded counterparts safe (albeit possibly slower than the original).

- **Forks:** We assume that $P$ does not fork in order to simplify the algorithms.
  If $P$ may fork with non-negligible probability, each sync phase could simply lag $k$ rounds behind the corresponding protocol phase. Adjusting $k$ such that the probability of overturning blocks after $k$ rounds is negligible.

### 6.1.2 Applicability to Existing Networks

To transform an active, non-sharded network into a sharded one, these steps could be followed:

1. Create a snapshot of the current state of the ledger (preferably as an account-balance list), formatted as a genesis block $B^0_{snapshot}$ .

2. Use $B^0_{snapshot}$ to bootstrap a sharded variant of the protocol.

## 6.2 Future Work

Our current model assumes that while nodes may join or leave arbitrarily, their order of magnitude remains constant throughout the entire execution. Supporting further dynamicity requires restructuring partitions. This might be required in case of massive departures, in order to ensure that honest majority is maintained. Also, in case of massive arrivals, when shards overcrowd it might be wise to adjust their number. The technical details of such scheme are left for future work.

Another possible extension is forming a "snitching" committee that collects proofs of Byzantine behavior in a specific shard and propagates them to participants outside that shard. Alternatively, per-shard validity proofs could be integrated into the protocol. These extensions can reduce the gap in tolerance to Byzantine nodes between 1/3 that is required by Algorand and 1/4 that we currently require, it may also enable dealing with a stronger adversary.

# Bibliography

[1] Saveen A Abeyratne and Radmehr P Monfared. Blockchain ready manufacturing supply chain using distributed ledger. 2016.

[2] Elli Androulaki, Artem Barger, Vita Bortnikov, Christian Cachin, Konstantinos Christidis, Angelo De Caro, David Enyeart, Christopher Ferris, Gennady Laventman, Yacov Manevich, Srinivasan Muralidharan, Chet Murthy, Binh Nguyen, Manish Sethi, Gari Singh, Keith Smith, Alessandro Sorniotti, Chrysoula Stathakopoulou, Marko Vukolic, Sharon Weed Cocco, and Jason Yellick. Hyperledger Fabric: A Distributed Operating System for Permissioned Blockchains. *CoRR*, abs/1801.10228, 2018.

[3] Rob A Bryer. Double-entry bookkeeping and the birth of capitalism: Accounting for the commercial revolution in medieval northern italy. *Critical perspectives on Accounting*, 4(2):113–140, 1993.

[4] Bruce G Carruthers and Wendy Nelson Espeland. Accounting for rationality: Double-entry bookkeeping and the rhetoric of economic rationality. *American journal of Sociology*, 97(1):31–69, 1991.

[5] Jing Chen, Sergey Gorbunov, Silvio Micali, and Georgios Vlachos. Algorand agreement: Super fast and partition resilient byzantine agreement. Cryptology ePrint Archive, Report 2018/377, 2018. https://eprint.iacr.org/2018/377.

[6] Ethereum. A Next-Generation Smart Contract and Decentralized Application Platform. 2018. URL https://github.com/ethereum/wiki/wiki.

[7] Yossi Gilad, Rotem Hemo, Silvio Micali, Georgios Vlachos, and Nickolai Zeldovich. Algorand: Scaling byzantine agreements for cryptocurrencies. In *Proceedings of the 26th Symposium on Operating Systems Principles*, SOSP '17, pages 51–68, New York, NY, USA, 2017. ACM. ISBN 978-1-4503-5085-3. doi: 10.1145/3132747.3132757. URL http://doi.acm.org/10.1145/3132747.3132757.

[8] Shafi Goldwasser and Rafail Ostrovsky. Invariant signatures and non-interactive zero-knowledge proofs are equivalent. In Ernest F. Brickell, editor, *Advances in Cryptology — CRYPTO' 92*, pages 228–245, Berlin, Heidelberg, 1993. Springer Berlin Heidelberg. ISBN 978-3-540-48071-6.

[9] Eleftherios Kokoris Kogias, Philipp Jovanovic, Nicolas Gailly, Ismail Khoffi, Linus Gasser, and Bryan Ford. Enhancing bitcoin security and performance with strong consistency via collective signing. In *25th {USENIX} Security Symposium ({USENIX} Security 16)*, pages 279–296, 2016.

[10] E. Kokoris-Kogias, P. Jovanovic, L. Gasser, N. Gailly, E. Syta, and B. Ford. OmniLedger: A Secure, Scale-Out, Decentralized Ledger via Sharding. In *IEEE Symposium on Security and Privacy (SP)*, pages 583–598, May 2018.

[11] Anna Lysyanskaya. Unique signatures and verifiable random functions from the dh-ddh separation. In Moti Yung, editor, *Advances in Cryptology — CRYPTO 2002*, pages 597–612, Berlin, Heidelberg, 2002. Springer Berlin Heidelberg. ISBN 978-3-540-45708-4.

[12] Bernard Marr. How Blockchain Will Transform The Supply Chain And Logistics Industry. mar. 2018.

[13] Petar Maymounkov and David Mazieres. Kademlia: A peer-to-peer information system based on the xor metric. In *International Workshop on Peer-to-Peer Systems*, pages 53–65. Springer, 2002.

[14] Silvio Micali. ALGORAND: the efficient and democratic ledger. *CoRR*, abs/1607.01341, 2016. URL http://arxiv.org/abs/1607.01341.

[15] Silvio Micali, Salil Vadhan, and Michael Rabin. Verifiable random functions. In *Proceedings of the 40th Annual Symposium on Foundations of Computer Science*, FOCS '99, pages 120–, Washington, DC, USA, 1999. IEEE Computer Society. ISBN 0-7695-0409-4. URL http://dl.acm.org/citation.cfm?id=795665.796482.

[16] Satoshi Nakamoto. Bitcoin: A peer-to-peer electronic cash system, 2009. URL http://www.bitcoin.org/bitcoin.pdf.

[17] David Schwartz, Noah Youngs, and Arthur Britto. The Ripple Protocol Consensus Algorithm. Ripple Labs Inc, 2014.

[18] Ewa Syta, Philipp Jovanovic, Eleftherios Kokoris Kogias, Nicolas Gailly, Linus Gasser, Ismail Khoffi, Michael J Fischer, and Bryan Ford. Scalable bias-resistant distributed randomness. In *2017 IEEE Symposium on Security and Privacy (SP)*, pages 444–460. Ieee, 2017.

[19] Mahdi Zamani, Mahnush Movahedi, and Mariana Raykova. Rapidchain: Scaling blockchain via full sharding. In *Proceedings of the 2018 ACM SIGSAC Conference on Computer and Communications Security*, CCS '18, pages 931–948, New York, NY, USA, 2018. ACM. ISBN 978-1-4503-5693-0. doi: 10.1145/3243734.3243853. URL http://doi.acm.org/10.1145/3243734.3243853.